\documentclass[acmsmall]{acmart}
\AtBeginDocument{%
  \providecommand\BibTeX{{%
    \normalfont B\kern-0.5em{\scshape i\kern-0.25em b}\kern-0.8em\TeX}}}

\setcopyright{acmcopyright}
\copyrightyear{2018}
\acmYear{2018}
\acmDOI{XXXXXXX.XXXXXXX}

\acmJournal{JACM}
\acmVolume{37}
\acmNumber{4}
\acmArticle{111}
\acmMonth{8}
\newcommand*{\revised}[1]{\textcolor{black}{#1}}

\usepackage{xcolor}

\usepackage{amsmath}
\usepackage{algorithm}
\usepackage{algpseudocode}

\usepackage{multirow}
\usepackage[normalem]{ulem}
\useunder{\uline}{\ul}{}
\usepackage{diagbox}

\usepackage{pifont}
\newcommand*{\revise}[1]{\textcolor{black}{#1}}

\usepackage{bbding}

\newcommand{\tool}{\textit{DSHGT}}

\algnewcommand\algorithmicoutput{\textbf{Output:}}
\algnewcommand\OUTPUT{\item[\algorithmicoutput]}

\algnewcommand\algorithmicdefine{\textbf{Define:}}
\algnewcommand\DEFINE{\item[\algorithmicdefine]}

\begin{document}

\title{\tool: Dual-Supervisors Heterogeneous Graph Transformer - A pioneer study of using heterogeneous graph learning for detecting software vulnerabilities}

\author{Tiehua Zhang}
\email{tiehuaz@tongji.edu.cn}
\thanks{Corresponding author: Tiehua Zhang}
\affiliation{
  \institution{Tongji University}
  \country{China}
}

\author{Rui Xu}
\email{17212010044@fudan.edu.cn}
\affiliation{
  \institution{Ping An Technology}
  \country{China}
}




\author{Jianping Zhang}
\email{17210240254@fudan.edu.cn}
\affiliation{
  \institution{Fudan University}
  \country{China}
}

\author{Yuze Liu}
\email{ericliu.lyz@gmail.com}
\affiliation{
  \institution{Ant Group}
  \country{China}
}

\author{Xin Chen}
\email{jinming.cx@antgroup.com}
\affiliation{
  \institution{Ant Group}
  \country{China}
}


\author{Jun Yin}
\email{jun.yinj@antgroup.com}
\affiliation{
  \institution{Ant Group}
  \country{China}
}

\author{Xi Zheng}
\email{james.zheng@mq.edu.au}
\affiliation{
  \institution{Macquarie University}
  \country{Australia}
}

\renewcommand{\shortauthors}{Zhang et al.}

\begin{abstract}
Vulnerability detection is a critical problem in software security and attracts growing attention both from academia and industry.
Traditionally, software security is safeguarded by designated rule-based detectors that heavily rely on empirical expertise, 
requiring tremendous effort from software experts to generate rule repositories for large code corpus. Recent advances in deep learning, 
especially Graph Neural Networks (GNN), have uncovered the feasibility of automatic detection of a wide range of software vulnerabilities. 
However, prior learning-based works only break programs down into a sequence of word tokens for extracting contextual features of codes, 
or apply GNN largely on homogeneous graph representation (e.g., AST) without discerning complex types of underlying program entities (e.g., methods, variables). 
In this work, we are one of the first to explore heterogeneous graph representation in the form of Code Property Graph 
and adapt a well-known heterogeneous graph network with a dual-supervisor structure for the corresponding graph learning task. 
Using the prototype built, we have conducted extensive experiments on both synthetic datasets and real-world projects. 
Compared with the state-of-the-art baselines, the results demonstrate superior performance in vulnerability detection (average F1 improvements over 10\% in real-world projects) and 
language-agnostic transferability from C/C++ to other programming languages (average F1 improvements over 11\%).
\end{abstract}

\ccsdesc[500]{Security and privacy~Software and application security}
\ccsdesc[300]{Software and its engineering~Software notations and tools}

\keywords{vulnerability detection, heterogeneous graph learning, code property graph (CPG)}


\maketitle
\pagestyle{plain}

\section{Introduction}
Software vulnerabilities are considered a major threat to system robustness and operability. 
The number of vulnerabilities reported and registered 
has been increasing significantly over the last decade owing 
to the growth of software practitioners and complex codebases ~\cite{lin2020software}. 
As a result, numerous methods and techniques have been developed to identify software vulnerabilities
especially at the early stage of development.

Many vulnerability detection tools have also been developed by big tech companies such as 
Meta (Getafix~\cite{bader2019getafix}), 
and Google (Tricorder~\cite{caitlin2015tricorder}). 
The underlying techniques can be broadly divided into two categories. 
Rule-based methods~\cite{ferschke2012flawfinder,viega2001static,kremenek2008finding, bader2019getafix, caitlin2015tricorder} 
take a set of expert-written rules to capture undesired code behaviours.
Learning-based methods~\cite{chowdhury2011using,shin2010evaluating,chernis2018machine}, on the other hand, 
intend to learn the underlying latent semantic and syntactic information and use the abnormal code corpus as the training samples. 
It has been shown in these studies (\textit{e.g.}~\cite{deepwukong},~\cite{lin} and~\cite{funded}) 
that the learning-based methods excel in detecting common code vulnerabilities or bugs than that of expert-crafted rules, 
especially with the recent advancements in deep learning techniques.

Many deep learning-based approaches model the source codes by capturing the shallow, contextual code token structures, 
which mainly use the recurrent neural network (RNN) and its variant language models~\cite{hanif2021rise}. 
These models are designed to split the code base and its abstract syntax tree into the sequence format, 
and thus ill-suited for encompassing the well-structured control dependency and data flows of programs. 
The use of graph neural networks (GNN) has recently emerged for solving code understanding tasks, 
owing to its potential to generalize both semantic and syntactic information. 
For instance, a gated GNN is first used in~\cite{allamanis2018learning} 
to represent the syntactic and semantic structure of a program snippet, 
which tends to solve the variable misuse and renaming problems rather than detecting general code vulnerabilities. 
Following that, a line of research started to explore the feasibility of using GNN to detect bugs in the programs. 

While it is tempting to unleash the power of GNNs to accomplish vulnerability detection tasks, 
encoding the code logic into a reasonable graph structure is non-trivial. 
Many works only expedite single relationships such as syntax hierarchy, data flow, 
and control dependency without concerning the heterogeneous attributes in the generated code graph, 
losing the generality of complex nodes and relation types in the graph~\cite{devign,deepwukong,funded}. 
The GNN models trained in this case are proven to perform undesirably in different tasks. 
Apart from that, another drawback of most GNN works is that the trained program model is constrained to only one type of programming language, 
thus failing to verify how transferable the model is in other programming languages. 
Also, the auxiliary information such as method-level \textcolor{black}{code comments} in the program often provides an extra dimension of code features 
which is rarely explored and could be helpful for improving the expressiveness of the model. 

In this paper, we conduct a pioneer study to explore whether using heterogeneous graph as the code representation 
and adapting a promising graph neural network for the representation learning 
can improve the performance of detecting software vulnerabilities especially across different language platforms and targeting real-world software projects.
For this purpose, we implement a novel heterogeneous graph learning-based framework, namely \tool.
\tool~uses and adapts a Heterogeneous Graph Transformer (HGT)~\cite{hu2020heterogeneous}, 
reporting a state-of-the-art performance on modelling the heterogeneous graph.
\tool~also uses Code Property Graph (CPG) to represent software programs, 
which was first proposed in~\cite{yamaguchi2014modeling} as the static vulnerability detection tool.
CPG merges elements of abstract syntax trees, control flow graphs and program dependence graphs into a joint graph structure. 
The rich intra-program relations and logical syntactic flows in the CPG serve iteself an ideal candidate for the heterogeneous graph representation in our study. 
Using \tool~for the intended heterogeneous graph learning, 
edge-driven weight matrices (e.g., for relationships) and node-driven attentions (e.g., for entities) 
derived from the initial CPG node embeddings can be parameterized specifically for the underlying heterogeneity. 
In such a way,
nodes and edges of different types in the CPG are able to maintain their specific representation, 
and \tool~is able to generate diverse embedding representations of the program suitable for the vulnerability detection task. 

Additionally, we leverage the annotation information in the code to enhance the embedding capability of \tool. 
The word token in the human-written \textcolor{black}{code comments} often contains supplementary semantic information 
about the program apart from code graph representations. 
To incorporate such information, \tool~introduces a multi-task learning mechanism, 
in which the trainable parameters in the model are updated by gradients with respect to 
both vulnerability detection loss and \textcolor{black}{code comment} generation loss, 
which we named dual-supervisors. In summary, the contributions of this paper are summarized as follows:

\begin{itemize}
\setlength{\itemindent}{-2.5em}

    \item[] \textbf{Pioneer Study:} We present a pioneer study of Heterogeneous Graph Learning for vulnerability detection by proposing and implementing \tool, 
    which embeds both semantic and heterogeneity properties of code representations (CPG) for improved vulnerability detection. 

    \item[] \textbf{Dual-supervisors Learning:}
    We design a multi-task learning framework with dual supervisors to utilize annotation information of codes to enhance the encoding capability of \tool, 
    which enables \tool~to generalize well to diverse programming languages and real-world software projects.

    \item[] \textbf{Extensive Experiments:} We conduct extensive experiments on both synthetic vulnerability datasets 
    across different programming languages and real-world projects to verify our hypothesis that using heterogeneous graph learning, 
    especially with dual-supervisor architecture, which can improve state-of-the-art software vulnerability detection and point out some interesting research directions for the community to follow. 

    \item[] \revise{\textbf{Source Code Release:} We have made the source code publicly available to contribute further to advancements in this field. The source code is available at https://github.com/Leesine/code2graph.}

\end{itemize}


The remainder of this paper is structured as follows. 
We first review the related work in Section~\ref{sec:related}. In Section~\ref{sec:pre}, we provide prerequisite backgrounds for our proposed \tool. In Section~\ref{sec:method}, we introduce the detailed methodology of \tool. We present the empirical study results and our discussion in Section~\ref{sec:exp}. Following that, we discuss the validity of our proposed method in Section~\ref{sec:validity} and draw the conclusion as well as future research direction in Section~\ref{sec:conclusion}.  

\section{Related Work}
\label{sec:related}

We take an overview of related works in software vulnerabilities detection from three different categories: 
traditional rule-based approach, deep learning-based approach, and graph learning-based approach. 

For the traditional approach, early works on vulnerability detection are heavily reliant on human-crafted rules from domain experts. 
The work in~\cite{engler2001bugs} is the first of this kind to implement rules to identify software bugs and vulnerabilities automatically. 
Following that, many static analysis tools~\cite{ferschke2012flawfinder,viega2001static,kremenek2008finding, bader2019getafix, caitlin2015tricorder} 
are developed to cover some well-known security issues, all of which share the same principle 
that if the scanned code base fails to conform to the re-defined rules, 
relevant vulnerabilities could occur. 
It is infeasible to craft rules that cover all possible code vulnerabilities, 
not to mention the required efforts to cope with the ever-changing code bases.

The rapid development of machine learning, especially deep learning techniques, 
unleashes the great potential in enabling the automated learning of implicit vulnerable programming patterns. 
Many early works focus on extracting the features from lines of codes to 
facilitate vulnerability detection/prediction~\cite{chowdhury2011using,shin2010evaluating,chernis2018machine}. 
For instance, 
VulDeePecker~\cite{li2018vuldeepecker} is the first deep learning-based binary vulnerability detector, 
which slices the program into code gadgets and utilizes BiLSTM to capture the semantic relations in the data dependence within the code gadgets. 
Similarly, $\mu$VulDeePecker~\cite{zou2019mu} uses both BiLSTM and code attentions to capture more ``localized'' information within a code statement, 
and control dependence among method calls. 
LIN \textit{et al.}~\cite{lin} designs a framework that uses data sources of different types for learning unified high-level representations of code snippets. 
It also uses BiLSTM as the core component of the learning process. 
DeepBugs~\cite{pradel2018deepbugs} uses a feedforward network as the classifier for name-based bug detection, 
which reasons about names based on semantic representations. 
However, only the natural code sequences are considered in these works, 
and the intra-program flow logic and dependency information are omitted.

Neural networks on graphs have drawn increasing attention in recent years, 
which focus on learning the model based on graph-structured input~\cite{li2015gated,kipf2016semi,zhang2021gps}. 
Researchers have put efforts into exploring the feasibility of using code graph representations such as 
Abstract Syntax Trees (AST), Program Dependency Graphs (PDG), and Control Flow Graphs (CFG) for vulnerability detection tasks. 
The work in~\cite{allamanis2018learning} first presents how to construct graphs from source code using AST and additional control and data flows, 
and then uses Gated Graph Neural Networks (GGNN) to detect variable misuse bugs. 
Afterwards, Devign~\cite{devign} utilizes four different code representations, including Abstract Syntax Tree (AST), Control Flow Graph (CFG), Data Flow Graph (DFG) and Natural Code Sequence (NCS), to construct the multi-relation code graph. Since Devign treats the multi-relation graph as multiple homogeneous graphs and applies Gated Graph Neural Network (GGNN) to propagate and aggregate the information among nodes in the code graphs, its capacity to encode heterogeneity is compromised compared with learning the heterogeneity attributions simultaneously. Apart from that, it substitutes data-related edge types with only three other relations (LastRead, LastWrite and ComputedFrom) in order to decrease the number of edge relations. Following
the work of Devign, AMPLE~\cite{wen2023vulnerability} constructs the code graph also based on AST, CFG, DFG, and NCS. It further simplifies the code graph through a set of manually defined rules, from which adjacent nodes and their edges are merged with the aim of reducing the total number of nodes and edges in the graphs. AMPLE then employs the graph convolutional network module to fuse heterogeneous information of the graph to update node embeddings. FUNDED~\cite{funded} integrates data and control flow into the AST as the extended-AST and starts to distinguish multiple code relationships when training the GGNN, which is achieved by representing the input program as multiple relation graphs. DeepWukong~\cite{deepwukong} combines the CFG and PDG to generate a refined subgraph called XFG for the program, and adopts three different GNNs to test the performance for bug prediction. MVD, on the other hand, only focuses on memory-related vulnerabilities by using a pruned code graph composed of a Program Dependency Graph (PDG) and a Call Graph (CG)~\cite{cao2022mvd}. It utilizes Flow-Sensitive Graph Neural Network (FS-GNN) to aggregate edge embeddings into the node embeddings by simply summing up the connected edges of different types, impairing MVD’s capacity to encode heterogeneity. DeepVD~\cite{wang2023deepvd} leverages three types of class-separation features at different levels of the code structure, including statement types, Exception Flow Graph (EFG) and Post-Dominator Tree (PDT), to which Label-GCN is applied to encode the code graph representation. However, DeepVD encompasses only the control flow edges that describe the execution order of code statements and lacks the data flow dependencies in the code graph.

Existing research mainly relies on adopting homogeneous graph learning techniques, from which types of nodes and edges are discarded, making it infeasible to represent heterogeneous structures. Moreover, it is proven in the domain of graph learning that integrating cross-relational interaction of diverse edge types simultaneously augments the capacity of models to encapsulate semantic and structural information within a heterogeneous graph~\cite{bing2023heterogeneous}.
We argue that the graph representations of codes convey rich semantic and logical information reflected 
in a variety of node/edge types and are intrinsic to the characteristics of heterogeneous graphs~\cite{wang2022survey}. This inspiration has propelled us to embark on an innovative investigation into the domain of heterogeneous graph learning, which, as will be demonstrated in subsequent sections, yields enhancements beyond the current state-of-the-art.

\section{Preliminary}
\label{sec:pre}

\begin{figure*}[t!]
\centering
\includegraphics[width=\textwidth]{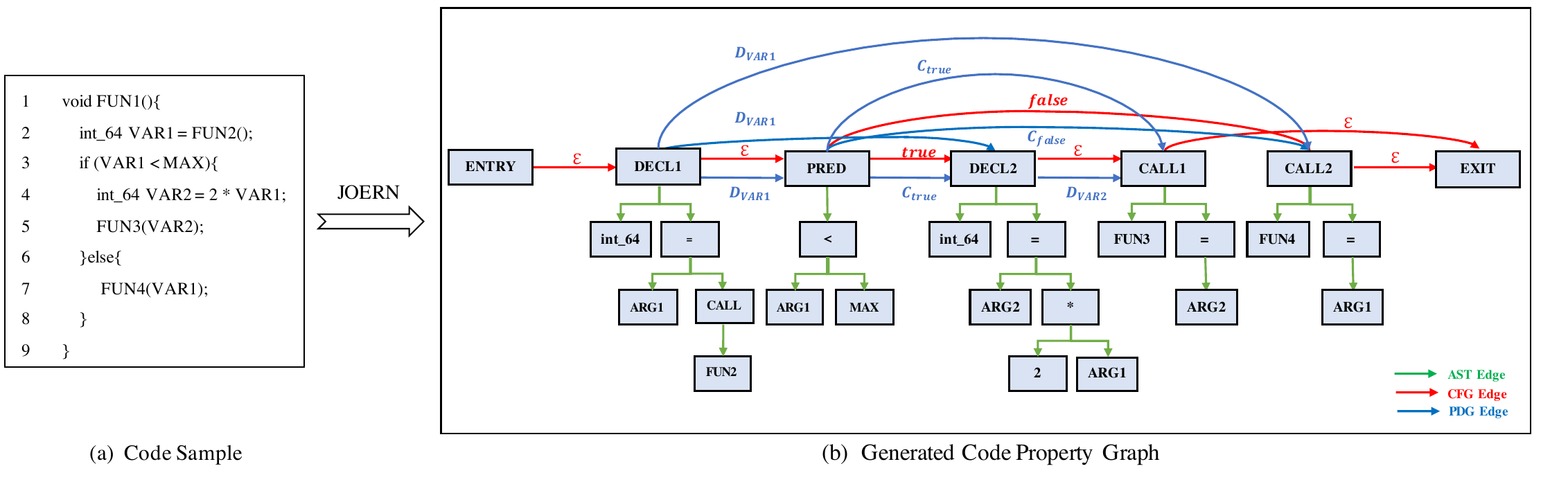}
\caption{Sample code and the generated CPG.}
\label{fig:cpg}
\end{figure*}

\subsection{Heterogeneous Graph}
\label{sec:hg}

In this section, we provide a formal definition of the heterogeneous graph~\cite{dong2017metapath2vec}. 
A heterogeneous graph $\mathcal{G = (V, E}, \mathcal{T}_v, \mathcal{T}_e)$ consists of 
a set of nodes $\mathcal{V}$ with multiple node types $\mathcal{T}_v$ and a set of edges $\mathcal{E}$ with multiple edge types $\mathcal{T}_e$. 
To describe the types of nodes and edges, we define two mapping functions: $ \tau : \mathcal{V} \rightarrow \mathcal{T}_v$ for the node types, 
and $ \phi : \mathcal{E} \rightarrow \mathcal{T}_e$ for the edge types. $\mathcal{X \in R^{N \times D}}$ represents the initial node feature matrix, 
where $\mathcal{N}$ is the number of nodes and $\mathcal{D}$ is the dimensionality of the feature vector. 
Finally, we define the Readout function $f : \mathcal{G} \rightarrow \mathcal{R}^{1 \times n}$, 
which summarizes the node and edge features and generates a low-dimensional graph-level embedding with $n$ dimensions.

\subsection{Code Property Graph}
The Code Property Graph (CPG) is a data structure representing source code, 
program structure, and execution logic in a unified graph-based representation~\cite{yamaguchi2014modeling}. 
It encapsulates the entire program in a single graph that consists of nodes and edges with multiple types, 
in which each node represents an entity, such as a function/method, a variable, or a class, 
and each edge represents a relationship between the entities. 
CPGs provide a comprehensive view of the program and enable the analysis of code properties 
and behaviors with the help of control flow, data flow, and intra-program dependence.

\textbf{Fig.~\ref{fig:cpg}} shows a toy example of CPG and its corresponding code snippet. 
For simplicity, we keep primary control/data flows, program dependence and execution steps for demonstrative purposes, 
while the specific AST edge types are not shown. 
The $\epsilon$ in \textbf{Fig.~\ref{fig:cpg}B} describes the control flow without any condition predicates when executing the program. 
The program steps over to the condition predicate statement and 
jumps to different statements based on \textit{true} or \textit{false} conditions (line 6 and 10 in \textbf{Fig.~\ref{fig:cpg}A)}. 
The blue edge represents both data and control dependence in the program. 
For instance, $D_{VAR1}$ shows the data dependence of variable \textit{VAR1} initially declared at line 3, 
and $C_{true}$ indicates that the variable declaration statement at line 6 is dependent on the condition predicate at line 5. 
The green edge is the simplified relationships in the abstract syntax tree (AST), 
where the inner nodes represent \textit{operators} (e.g., addition, assignment) 
and leaf nodes as \textit{operands} (e.g., variable identifier, constants). 
The full descriptions of node types and edge types in the CPG can be found in Table~\ref{tab:node} and Table~\ref{tab:edge}.

\begin{figure*}[t!]
\centering
\includegraphics[width=0.98\textwidth]{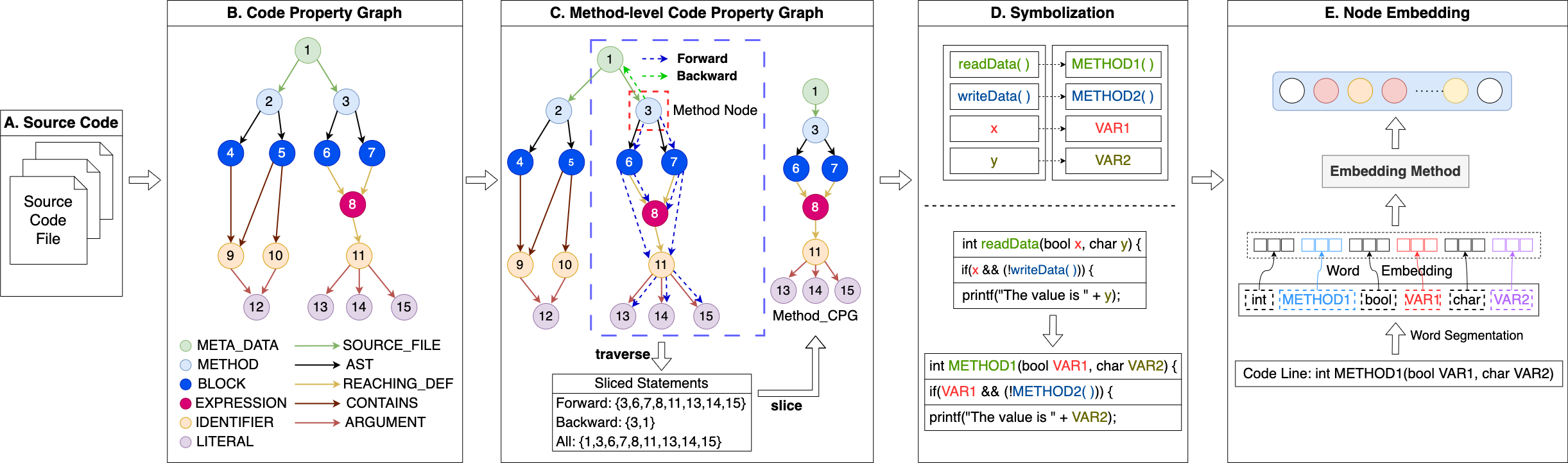}
\caption{Graph Construction}
\label{fig:graph-cons}
\end{figure*}

\subsection{Heterogeneous Graph Transformer}
\label{sec:hgt}

To capture the complex interactions between various types of entities in heterogeneous graphs, 
Hu \textit{et al.} proposed Heterogeneous Graph Transformer (HGT)~\cite{hu2020heterogeneous}, 
which introduces the transformer into the graph neural network to better incorporate the heterogeneous nature of the graph structure.

In HGT,
a triplet $<\tau(s), \phi(e) ,\tau(t)>$ can be used to denote the relationships among 
source node ($s$), directed edge ($e$), and target node ($t$).
The function $\tau(\cdot)$ is node type mapping, which outputs the type of input node. 
Similarly, function $\phi(\cdot)$ denotes the edge type mapping, which outputs the type of input edge. 
This triplet and the original node embedding for the source and target nodes are the input for HGT. 
The embedding of nodes is updated through multiple iterations, each of which includes three steps:
\textit{1)} Attention calculation (Section~\ref{sec:attention});
\textit{2)} Message calculation (Section~\ref{sec:message});
and \textit{3)} Aggregation of attention and message (Section~\ref{sec:aggregate}). 
We will detail our adaptation of HGT in Section~\ref{sec:method-hgt-multi-task}.

\section{Heterogeneous Graph Learning}
\label{sec:method}

In our proposed heterogeneous graph learning procedure, we have the following two main steps. 
The first step is Graph Construction (shown in \textbf{Fig.~\ref{fig:graph-cons}}). 
We analyze all the source code and generate the initial CPG
(\textbf{Fig.~\ref{fig:graph-cons}A} \& \textbf{Fig.~\ref{fig:graph-cons}B}). 
As we focus on method-level vulnerability analysis in this pioneering study, 
we extract method-level CPGs from the initial CPG  (\textbf{Fig.~\ref{fig:graph-cons}C}).
Meanwhile, symbolization is also performed on method-level CPGs to 
reduce the noise introduced by personalized function/variable naming conventions (\textbf{Fig.~\ref{fig:graph-cons}D}).
We then perform the embedding method for each node within 
the method-level CPGs for the next step (\textbf{Fig.~\ref{fig:graph-cons}E}).

The second step is our adaptation of HGT - Dual-Supervisors HGT Learning (shown in \textbf{Fig.~\ref{fig:hgt-multi}}). 
In Dual-Supervisors  HGT learning, 
we use initial node features as the input of HGT to learn and extract the graph-level information. 
\textcolor{black}{HGT can effectively encode the heterogeneity of CPG, which helps improve the generalization ability of the model.}
We then leverage dual-supervisors learning for both vulnerability prediction and \textcolor{black}{code comment} generation. 
We introduce the \textcolor{black}{code comment} as the second supervisor to align the latent features 
learned from the HGT with the underlying semantic meaning through the back-propagation process.

\subsection{Graph Construction}
\label{sec:method-graph}

We first analyze all the source code files 
and generate the initial corresponding Code Property Graphs (CPGs). 
In our case, all the related files (\textit{e.g.} source code files and dependency library files) 
are within the same directory. By inputting this directory to an open source 
code parser \textit{Joern}\footnote{https://github.com/joernio/joern}, 
these source code files are then iterated automatically to generate the corresponding CPG.

\begin{algorithm}[t!]
\caption{Generating Method-level Code Property Graph}
\label{alg:method-cpg}
\begin{algorithmic}[1]

\Require Source code root directory \textit{S}, \textit{Joern} parser \textit{J}
\OUTPUT Method-level CPGs set \textit{M}.
\State \textit{C} $\leftarrow$ $\emptyset$
\For{each leaf directory \textit{l} $\in$ dir(\textit{S})}
    \State  generate CPG \textit{c} through \textit{J}(\textit{l}), and add to set \textit{C}
\EndFor
\State \textit{M} $\leftarrow$ $\emptyset$
\For{each \textit{c} $\in$ \textit{C}}
    \State  \textit{N} $\leftarrow$all \textit{method} type nodes within \textit{c}
    \For{each \textit{n} $\in$ \textit{N}}
        \State start at \textit{n}, perform DFS forward traverse  
        \State start at \textit{n}, perform DFS backward traverse
        \State generate method-level CPG \textit{m} for method \textit{n}
        \State add \textit{m} to set \textit{M}
    \EndFor
\EndFor

\end{algorithmic}
\end{algorithm}

In this paper, we concentrate on method-level vulnerability analysis, 
and Algorithm~\ref{alg:method-cpg} demonstrates how we construct the method-level CPGs. 
As indicated in the pseudo-code,
the generated CPG is denoted as $\mathit{c}$, which contains all relationships of source codes
within one leaf directory. The set of all directory-level CPGs is denoted as $\mathit{C}$. 
Instead of using original CPG $\mathit{c}$ that contains much redundant information in the graph, 
we perform forward and backward traversal to generate the method-level CPG $\mathit{m}$. 
Specifically, both traversals are based on Depth-First Search (DFS) for each \textit{Method} node within $\mathit{c}$, 
and the set of all method-level CPGs is denoted as $\mathit{M}$. Taking \textbf{Fig~\ref{fig:graph-cons}C} as an example, 
node \textit{3} is a method node, from which we traverse forward through nodes \textit{6,7,8,11,13,14,15}, 
while traversing backward through the node \textit{1}. 
Thus all the nodes be traversed are \textit{1,3,6,7,8,11,13,14,15} including itself.
Thus, the corresponding method-level CPG (Method\_CPG) could be generated by slicing this traversed set out of the original CPG.

In each CPG (\textbf{Fig~\ref{fig:graph-cons}B}), 
we construct the heterogeneous graph by mapping the original entities (e.g., Method name) and relationships (e.g., method call) to
different types of nodes (e.g., \textit{METHOD}) and edges (e.g., \textit{CALL}).
For a full list of nodes and edges we generated for CPG, 
please refer to Table~\ref{tab:node} and Table~\ref{tab:edge}.

Meanwhile, 
to alleviate the noise introduced by personalized naming conventions 
for functions and variables and better preserve the original code semantics~\cite{deepwukong}, 
we then perform symbolization on method-level CPGs (shown in \textbf{Fig~\ref{fig:graph-cons}D}). Following that,
different function and variable names defined by users will be unified to \textit{METHOD`N'(~)} and \textit{VAR`N'}, where $N\in\mathbb{Z}^{+}$. 
For example, the function names \textit{readData(~)} and \textit{writeData(~)} 
and variable names \textit{x} and \textit{y} 
will be mapped to \textit{METHOD1(~)}, \textit{METHOD2(~)}, \textit{VAR1}, and \textit{VAR2}, respectively. 
The actual numbers $N$ used in the symbolization may vary.
As shown in \textbf{Fig~\ref{fig:graph-cons}E},
we then perform Doc2Vec embedding~\cite{le2014distributed} for each node within the method-level CPGs. 
This embedding serves as the initial node feature and will be refined during the Dual-Supervisors HGT learning.

\subsection{Dual-Supervisors HGT Learning}
\label{sec:method-hgt-multi-task}

\begin{figure*}[t!]
\centering
\includegraphics[width=\textwidth]{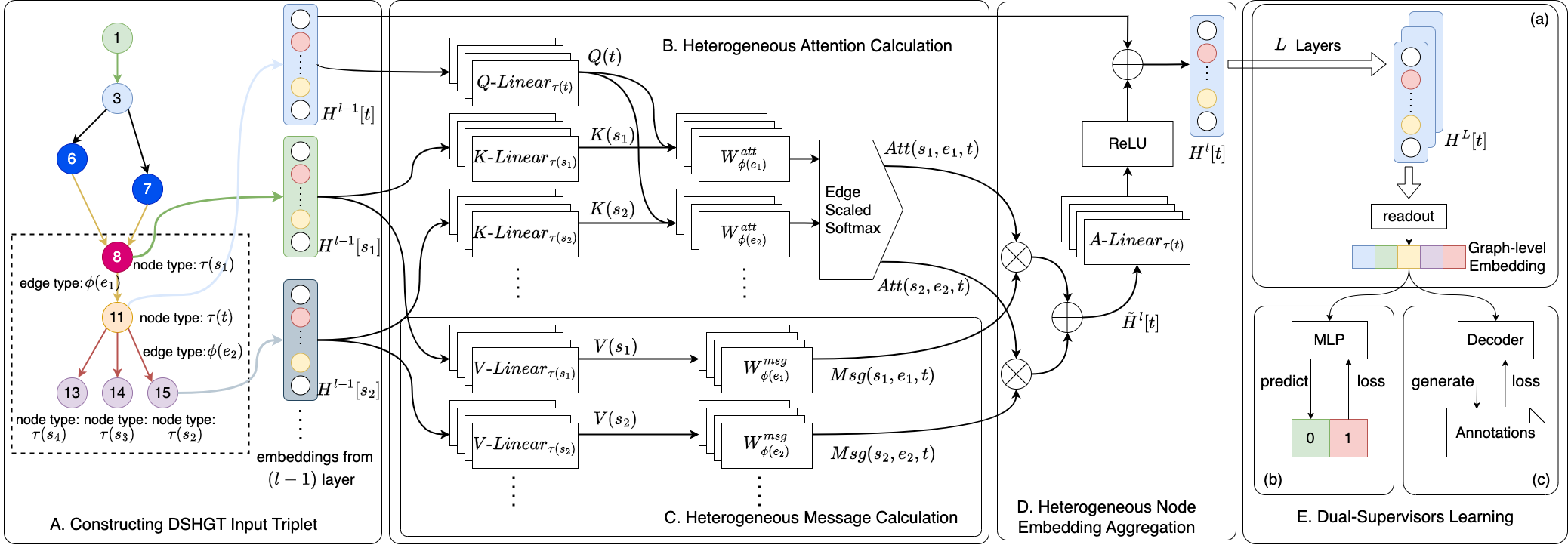}
\caption{\textit{DSHGT} Learning}
\label{fig:hgt-multi}
\end{figure*}

The overall architecture of the Dual-Supervisors HGT (\textit{DSHGT}) is shown in \textbf{Fig~\ref{fig:hgt-multi}}.
For each target node in a given CPG, we consider all its connected neighbors as source nodes, 
and for each source node/target node pair we define $<\tau(s),\phi(e),\tau(t)>$ triplet as the relationship of this pair (shown in \textbf{Fig~\ref{fig:hgt-multi}A}). 
For each triplet, we then calculate the attention score between the source node and the target node (shown in \textbf{Fig~\ref{fig:hgt-multi}B}), 
calculate the messaging passing score of each source node (shown in \textbf{Fig~\ref{fig:hgt-multi}C}), 
and aggregate the information from the above two steps and update the target node embedding (shown in \textbf{Fig~\ref{fig:hgt-multi}D}). 
To improve the robustness of the learned features (aggregated target node embedding), 
we use the existing \textcolor{black}{code comment} for each method as the additional supervisor in the multi-task learning (shown in \textbf{Fig~\ref{fig:hgt-multi}E}). 
We will walk through each step in more details below.

Note, we use $H^{\alpha}[\beta]$ to denote node $\beta$'s embedding in the $\alpha$-th layer, 
and $H^{\alpha-1}[\beta]$ to denote node $\beta$'s embedding in the $(\alpha-1)$-th layer, 
through the whole section.

\subsubsection{Constructing \textit{DSHGT} Input Triplet (Fig~\ref{fig:hgt-multi}A)} 

For each method CPG, we iteratively walk through the CPG using the depth-first search algorithm and construct the triplet from the root level all the way to the leaf nodes. 
For instance, when we walk to the node of \{11\} (\textit{i.e.} node $t$) in \textbf{Fig~\ref{fig:hgt-multi}A}, 
we treat the node as the current target node. 
We found its neighbor nodes are \{8\}, \{15\}, \{14\} and \{13\} (\textit{i.e.} node $s_1$, $s_2$, $s_3$ and $s_4$, respectively). 
Then we construct the following triplets 
$<\tau(s_1), \phi(e) ,\tau(t)>$, 
$<\tau(s_2), \phi(e) ,\tau(t)>$, 
$<\tau(s_3), \phi(e) ,\tau(t)>$ 
and 
$<\tau(s_4), \phi(e) ,\tau(t)>$.
We then feed them all to the \textit{DSHGT}. 
Note, to simplify the figure, we only present the embedding 
for node $t$, $s_1$ and $s_2$ in \textbf{Fig~\ref{fig:hgt-multi}}.

\subsubsection{Heterogeneous Attention Calculation (Fig~\ref{fig:hgt-multi}B)}
\label{sec:attention}

Firstly, we calculate the attention between $s_1$ and $t$, where $s_1$ is one of the neighbor nodes of $t$. 
The calculation involves five equations (Eq.~\ref{eq:q-linear} to Eq.~\ref{eq:attention}).

\begin{equation}
\label{eq:q-linear}
    Q^i(t) =   H^{(l-1)}[t] \cdot Q{\text{-}}Linear^i_{\tau(t)}
\end{equation}
where the dimension of $H^{(l-1)}[t]$ (\textit{i.e.} the embedding of $t$) is $R^{1 \times d}$, 
the dimension of $Q{\text{-}}Linear^i_{\tau(t)}$ is $R^{d \times \frac{d}{h}}$, 
 $i$ ($i \in [1,h]$) represent the $i$-th head of attention, 
and the dimension of $Q^i(t)$ is $R^{1 \times \frac{d}{h}}$.
We project $H^{(l-1)}[t]$ to $Q^i(t)$ through $Q{\text{-}}Linear^i_{\tau(t)}$ (\textit{i.e.} the query matrix), 
and each $\tau(\cdot)$ has 
its own matrix $Q{\text{-}}Linear^i_{\tau(t)}$ on the $i$-th head.

\begin{equation}
\label{eq:k-linear}
 K^i(s_1) =  H^{(l-1)}[s_1] \cdot K{\text{-}}Linear^i_{\tau(s_1)}
 \end{equation}

In Eq.~\ref{eq:k-linear}, the dimension of $H^{(l-1)}[s_1]$ (\textit{i.e.} the embedding of $s_1$) is $R^{1 \times d}$, 
the dimension of $K{\text{-}}Linear^i_{\tau(s_1)}$ is $R^{d \times \frac{d}{h}}$, 
where $i$ ($i \in [1,h]$) represent the $i$-th head of attention, 
and the dimension of $K^i(s_1)$ is $R^{1 \times \frac{d}{h}}$.
We project $H^{(l-1)}[s_1]$ to $K^i(t)$ through $K{\text{-}}Linear^i_{\tau(s_1)}$ (\textit{i.e.} the key matrix), 
and each $\tau(\cdot)$ has 
its own matrix $K{\text{-}}Linear^i_{\tau(s_1)}$ on the $i$-th head.

\begin{equation}
\label{eq:i-attention}
ATT{\text -}head^i(s_1,e,t)=(K^i(s_1) \times W^{ATT}_{\phi(e)} \times Q^i(t)^T) \times \frac{\mu_{<\tau(s_1),\phi(e),\tau(t)>}}{\sqrt{d}} 
\end{equation}

Eq.~\ref{eq:i-attention} is for calculating the attention value of the $i$-th head from $s_1$ to $t$.
The $W^{ATT}_{\phi(e)} \in R^{\frac{d}{h} \times \frac{d}{h}}$ stands for a learnable parameter matrix for edge type $\phi(e)$,
which represents the learnable semantic information for each edge type. 
The $(K^i(s_1) \times W^{ATT}_{\phi(e)} \times Q^i(t)^T)$ is the original attention value of the $i$-th head. 
The dimension of it is $R^{1 \times 1}$ 
(\textit{i.e.} $R^{1 \times \frac{d}{h}} \times R^{\frac{d}{h} \times \frac{d}{h}} \times R^{\frac{d}{h} \times 1} \rightarrow R^{1 \times 1}$). 
The $\mu$ is a matrix related with the triplet $<\tau(s_1),\phi(e),\tau(t)>$, 
which acts as a scaling factor for this triplet relationship.
Its dimension is $R^{A \times E \times A}$, 
where $A = |\tau(\cdot)|$ and $E = |\phi(\cdot)|$. 
It is worth noting that, the magnitude of $K$ and $Q$ dot product increases significantly
and will lead \textit{Softmax} function to small gradient values.
Thus, we divide the original value by $\sqrt{d}$ to
maintain the gradient values after \textit{Softmax}, 
which could help the training.

\begin{equation}
\label{eq:1-attention}
\textbf{Attention}_{DSHGT}(s_1, e ,t) = \mathop{\parallel}\limits_{i \in[1, h]} ATT{\text -}head^i(s_1, e, t)
\end{equation}

Eq.~\ref{eq:1-attention} is for calculating the attention value from $s_1$ to $t$. 
The $h$ heads attention values from Eq.~\ref{eq:i-attention} will be concatenated 
together to a vector with $R^{h \times 1}$ dimensions.
Note the attention calculation will be the same 
for all the triplets, $<\tau(s_2), \phi(e) ,\tau(t)>$, $<\tau(s_3), \phi(e) ,\tau(t)>$, \textit{etc.}

To yield the final attention value for each head, we gather attention matrices 
from all neighbors of $N(t)$ and conduct \textit{Softmax} as shown in Eq.~\ref{eq:attention}. 

\begin{equation}
\label{eq:attention}
\textbf{Attention}_{DSHGT}(s, e ,t) = \mathop{Softmax}\limits_{\forall{s \in N(t)}} \left(   \textbf{Attention}_{DSHGT}(s, e ,t)   \right)
\end{equation}

\subsubsection{Heterogeneous Message Calculation (Fig~\ref{fig:hgt-multi}C)}
\label{sec:message}

Secondly, we shows how to calculate the message from $s_1$ to $t$, 
which involves three equations (Eq.~\ref{eq:v-linear} to Eq.~\ref{eq:message}).

\begin{equation}
\label{eq:v-linear}
V^i(s_1) = H^{(l-1)}[s_1] \cdot V{\text{-}}Linear^i_{\tau(s_1)}
\end{equation}

In Eq.~\ref{eq:v-linear}, 
the dimension of $H^{(l-1)}[s_1]$ is $R^{1 \times d}$, 
the dimension of $V{\text{-}}Linear^i_{\tau(s_1)}$ is $R^{d \times \frac{d}{h}}$, 
where $i$ ($i \in [1,h]$) represent the $i$-th head of message, 
and the dimension of $V^i(s_1)$ is $R^{1 \times \frac{d}{h}}$.
We project $H^{(l-1)}[s_1]$ to $V^i(s_1)$ through $V{\text{-}}Linear^i_{\tau(s_1)}$ (\textit{i.e.} the value dimension), 
and each $\tau(\cdot)$ has 
its own parameter $V{\text{-}}Linear^i_{\tau(s_1)}$ on the $i$-th head.

\begin{equation}
\label{eq:i-message}
    MSG{\text -}head^i(s_1, e, t) = V^i(s_1) \times W^{MSG}_{\phi(e)} 
\end{equation}

Eq.~\ref{eq:i-message} calculates the message value of the $i$-th head from $s_1$ to $t$.
The $W^{MSG}_{\phi(e)} \in R^{\frac{d}{h} \times \frac{d}{h}}$ stands for a learnable parameter matrix for edge type $\phi(e)$, 
and each $\phi(\cdot)$ has its own $W^{MSG}_{\phi(e)}$ matrix. 
The dimension of $MSG{\text -}head^i(s_1, e, t)$  is $R^{1 \times \frac{d}{h}}$ 
(\textit{i.e.} $R^{1 \times \frac{d}{h}} \times R^{\frac{d}{h} \times \frac{d}{h}} \rightarrow R^{1 \times \frac{d}{h}} $).

\begin{equation}
\label{eq:message}
    \textbf{Message}_{DSHGT}(s_1, e ,t) = \mathop{\parallel}\limits_{i \in[1, h]} MSG{\text -}head^i(s_1, e, t)
\end{equation}

The message value from $s_1$ to $t$ is calculated in Eq.~\ref{eq:message}, 
in which the $h$ heads message values from Eq.~\ref{eq:i-message} will be concatenated 
together to a matrix with $R^{h \times \frac{d}{h}}$ dimensions.
Note the message calculation will be the same 
for all the triplets, $<\tau(s_2), \phi(e) ,\tau(t)>$, $<\tau(s_3), \phi(e) ,\tau(t)>$, \textit{etc.}

\subsubsection{Heterogeneous Node Embedding Aggregation(Fig~\ref{fig:hgt-multi}D)}
\label{sec:aggregate}

Thirdly, we calculate the aggregation of attention and message from $s_1$ to $t$. 

\begin{equation}
\label{eq:aggregate-1}
    \tilde{H}^l[t_{s_1}] =  \textbf{Attention}_{DSHGT}(s_1, e ,t) \otimes \textbf{Message}_{DSHGT}(s_1, e ,t)
\end{equation}

The weighted message from $s_1$ is shown in Eq.~\ref{eq:aggregate-1},
where $\otimes$ is element-wise multiplication. The dimension of $\textbf{Attention}_{DSHGT}(s_1, e ,t))$ is $R^{h \times 1}$, 
and the dimension of $\textbf{Message}_{DSHGT}(s_1, e ,t)$ is $R^{h \times \frac{d}{h}}$. 
Note the weighted message calculation will be the same for all the triplets.

After calculating the weighted message for all neighbors of $N(t)$, 
we could update the target node $t$ embedding based on messages from its neighbors. 

\begin{equation}
\label{eq:aggregate-2}
    \tilde{H}^l[t] = \mathop{\oplus}\limits_{\forall{s \in N(t)}} \tilde{H}^l[t_{s}]
\end{equation}

We then reshape the vector to $\tilde{H}^l[t] \in R^{1 \times d}$.

\begin{equation}
\label{eq:aggregate-residual}
    H^l[t] = \alpha\left( \tilde{H}^l[t] \cdot A{\text -}Linear_{\tau(t)} \right) + H^{l-1}[t] 
\end{equation}

In Eq.~\ref{eq:aggregate-residual},
the $\tilde{H}^l[t]$ stands for the node $t$ embedding for current layer, 
the $H^{l-1}[t]$ stands for the node $t$ embedding from previous layer, 
the $\alpha$ is the activation function (\textit{i.e.} ReLU). 
We project $\tilde{H}^l[t]$ through $A{\text{-}}Linear_{\tau(t)} \in R^{d \times d}$. 
Note each $\tau(\cdot)$ has its own parameter in $A{\text{-}}Linear_{\tau(t)}$. 
The projection (\textit{i.e.} $A{\text -}Linear_{\tau(t)}\left(\tilde{H}^l[t]\right)$) will then 
go through the activation function before adding up the node $t$ embedding from the previous layer as residual, 
and yield the final node $t$ embedding for the current layer.
In one iteration, we update all the nodes' embedding within the heterogeneous graph following the same procedure.
We iterate this calculation for every nodes within the method-level CPG for $L$ layers. 
The $L$-th layer output $H^L[t]$ (\textbf{Fig~\ref{fig:hgt-multi}D}) will be used for downstream tasks.

\subsubsection{Dual-Supervisors Learning (Fig~\ref{fig:hgt-multi}E)}

The \textit{DSHGT} node embedding procedure will go through $L$ times (\textit{i.e.} $L$ layers of \textit{DSHGT}),
and each layer will use the previous layer's embedding as input 
(the initial layer's input is based on the CPG embedding, 
details in Section~\ref{sec:method-graph}).
In our experiments, we perform empirical study and set $L = 3$ (details analyzed in Section~\ref{sec:rq3}).
As the output form \textit{DSHGT} (\textit{i.e.} $H^L[t]$) is node-based embedding, 
we construct a \textit{Readout} layer for graph-level embedding output:

\begin{equation}
\label{readout}
    z^{\mathcal{G}}=M E A N\left(M L P\left(\mathcal{X} \oplus H^L\right)\right)
\end{equation}

Instead of directly taking out the embeddings, 
we concatenate them with the initial node embedding, pass through a shallow multi-layer perceptron (MLP),
and follow with a row-wise \textit{MEAN} operation.
$\mathcal{X}$ is defined in Section~\ref{sec:hg}, which represents the initial node embedding. 

This output then goes through a dual-supervisors structure(\textit{i.e.} MLP and Decoder) 
for multi-task purposes (\textbf{Fig~\ref{fig:hgt-multi}E}).
For detecting vulnerabilities within the source code (\textbf{Fig~\ref{fig:hgt-multi}E(\textit{b})}), 
we use 1 layer MLP for \textbf{0/1} classification, 
where \textbf{0} stands for no vulnerability while \textbf{1} means the source code segment contains vulnerabilities.
\textcolor{black}{We only test whether the code snippets contain vulnerabilities, and do not classify them into a specific type of CWE.}
On the other hand, We consider the graph-level embedding as an \textit{Encoder} for the source code 
and design the corresponding \textit{Decoder} (1-layer LSTM) to summarize 
the corresponding source \textcolor{black}{code comments} sequence-to-sequence (\textbf{Fig~\ref{fig:hgt-multi}E(\textit{c})}). 
Then we compare the generated \textcolor{black}{code comments} with \textcolor{black}{comments} within source code (\textit{i.e.} the ground truth) 
and yield cross entropy loss for multi-tasks. 
To leverage the \textit{loss} from the two supervisors, 
we implement the following equation for \textit{loss} fusion: 

\begin{equation}
\label{eq:multi-task-weight}
    loss = (1-\lambda) \times loss_{main} + \lambda \times loss_{sup}
\end{equation}

\revised{In Eq.~\ref{eq:multi-task-weight}, 
$loss_{main}$ is the \textit{loss} of $0/1$ classification and $loss_{sup}$ is the \textit{loss} of code comments prediction. 
The $\lambda$ is the parameter for adjusting the weight of $loss_{sup}$ in $loss$.}

\section{Experiment}
\label{sec:exp}

We evaluate the performance of our framework on different datasets 
against a number of state-of-the-art graph-based or traditional vulnerability detection models. 
We aim to answer the following research questions:

\begin{itemize}
\setlength{\itemindent}{-2.5em}
    \item[] \textbf{RQ1}: How well our proposed framework performs compared with other baselines in public C/C++ vulnerability datasets?\par
    \textbf{Motivation}: This RQ aims to assess the performance of our framework against existing state-of-the-art methods (as baselines) in public C/C++ vulnerability datasets. We conduct experiments on the most representative 22 vulnerability categories for comprehensive performance comparison and analysis through widely-used evaluation metrics.
    \item[] \textbf{RQ2}: Can the framework achieve a consistently higher vulnerability detection capability when applied to other programming languages? \par
    \textbf{Motivation}: This RQ investigates the cross-language vulnerability detection capabilities of DSHGT. It aims to answer whether DSHGT can detect similar vulnerabilities in other programming languages (i.e., PHP, Java, etc.) if only trained in one programming language (C/C++), from which the transferability of the proposed framework can be well studied.
    \item[] \textbf{RQ3}: How to balance the contribution from the two supervisors 
    (i.e., vulnerability and \textcolor{black}{code comment} oracles) to improve the performance?\par
    \textbf{Motivation}: We are motivated to investigate the necessity of designing multi-task learning with dual supervisors in DSHGT for dealing with the vulnerability label oracle and the code comment oracle. Additionally, we aim to study how much the code comments, as supplementary information, can contribute to the overall performance when detecting the vulnerabilities.
    \item[] \textbf{RQ4}: How much can the CPG input representation and HGT backbone improve the performance?\par
    \textbf{Motivation}: This RQ aims to evaluate the impact of different code graph representations on heterogeneous graph learning using HGT. The significance of using CPG and its heterogeneous edge types, which encompasses rich semantic and syntactic data of code snippets, is unveiled through a detailed ablation study, including the comparison with different code graph representations such as AST, PDG, AST+CFG, and AST+PDG. We then study whether learning the heterogeneity attributes in these code graph representations can help improve the performance in vulnerability detection tasks.
    \item[] \textbf{RQ5}: How effective is our proposed method when applied to detect vulnerabilities in real-world open-source projects? \par
    \textbf{Motivation}: We are motivated to further validate the practicality of DSHGT in detecting vulnerabilities in real-world projects. 
\end{itemize}

\revise{The rationale for designing these research questions are: 1) To evaluate the comprehensive efficacy of our proposed framework, DSHGT, in performing vulnerability detection tasks on the extensively utilized Software Assurance Reference Dataset (SARD), and to benchmark it against four prevailing state-of-the-art baseline models (RQ1); 2) To ascertain whether DSHGT demonstrates a programming language-agnostic ability for vulnerability detection, such that training the model on one programming language allows it to retain knowledge of vulnerable code patterns, which can then be applied to identify similar vulnerabilities across different programming languages (RQ2); 3) To investigate the impact of incorporating code comments on enhancing the understanding of code semantics, due to the additional context they provide, and to assess the potential of bolstering vulnerability detection performance by encoding comment information authored by domain experts (RQ3); 4) To examine the significance of employing the Code Property Graph (CPG) in the domain of vulnerability detection and to explore whether leveraging the heterogeneity attributes within the graph structure can amplify the performance in identifying vulnerabilities (RQ4); 5) To probe the practical applicability of DSHGT in the context of unearthing vulnerabilities within more intricate, real-world software systems (RQ5).}

\subsection{Experimental Setup}
We describe the experimental setup in this section, 
including the environment, baselines, evaluation metrics, and the preparation of the dataset.
\textcolor{black}{The parameter statistics of models are shown in Table \ref{tab:model}.}

\begin{table}[h!]
\centering
\caption{\textcolor{black}{\textcolor{black}{Information of models parameters}}}
\label{tab:model}
\scalebox{0.9}{
\begin{tabular}{|c|c|c|c|c|}
\hline
Model Information & HGT     & Decoder & MLP    & Total   \\ \hline
Number of Parameters  & 5,091,894 & 46,280   & 12,456  & 5,150,720 \\ \hline
Size              & 5.09 MB & 0.05 MB & 0.01MB & 5.15MB  \\ \hline
\end{tabular}
}
\end{table}

\begin{table}[t!]
\centering
\caption{Hyperparameter setup}
\label{tab:hyper}
\scalebox{0.8}{
\begin{tabular}{c||c}
\hline
Name & \ Setup\\
\hline
\hline
Readout func (HGT) & 2 linear layers, 1 output layer   \\
Layer depth (HGT) & 3 \\
Attention head (HGT) & 4 \\
Loss function & Cross entropy loss   \\
Optimizer & Adam~\cite{kingma2014adam}\\
Learning rate & 2e-3 \\
Dropout rate & 0.5 \\
Batch size & 64\\
Epochs & 50\\
Weight initializer & Xavier~\cite{glorot2010understanding}\\

\hline
\end{tabular}
}
\end{table}

\subsubsection{Environment}
We implemented our heterogeneous graph-based vulnerability detection model using Python v3.7 and Pytorch v1.11.0. 
As mentioned in Section~\ref{sec:method-graph}, 
we leveraged \textit{Joern} to 
generate the initial CPG of different programming languages. 
We trained and tested our model on a computer with an 8-core 3.8 GHz Intel Xeon CPU and an NVIDIA 3080Ti GPU. 
The hyperparameter setup could be found in Table~\ref{tab:hyper}.

\subsubsection{Baselines}
We compared our \tool~with the following state-of-the-art baselines and reported the comparison statistics.

\begin{itemize}
\setlength{\itemindent}{-2.5em}

    \item[] \textbf{LIN \textit{et al.}}~\cite{lin} designs a framework that uses data sources of 
    different types for learning unified high-level representations of code snippets. 
    It uses BiLSTM as the core component of the learning process.

    \item[] \textbf{DEVIGN}~\cite{devign} combines AST, program control and data 
    dependency as the joint graph to represent the composite code representation, 
    from which a gated graph neural network model is designed 
    to learn the graph-level embeddings for the vulnerability detection task.
    
    \item[] \textbf{FUNDED}~\cite{funded} integrates data and control flow into the AST as the code graph representation, 
    which is then used as the input for the gated graph neural network (GGNN) to train the vulnerability detection model. 
    It uses Word2Vec~\cite{mikolov2013efficient} to generate the initial node embedding. 
    
    \item[] \textbf{DeepWukong}~\cite{deepwukong} is also a graph learning-based approach that 
    encodes both textual and structured information of code into code representations. 
    It is designed specifically to detect C/C++ vulnerabilities. 
    It uses Doc2Vec~\cite{le2014distributed} to generate initial node embeddings from PDG.
    
\end{itemize}

Note that for all baseline models, we used the default hyperparameters as reported in the respective literature.

\subsubsection{Evaluation Metrics}

We used \textbf{Accuracy}, \textbf{Precision}, \textbf{Recall} 
and \textbf{F1} scores to evaluate the vulnerabilities detected by a model, 
which are widely used in the machine learning community to 
verify the generalization ability of a predictive model~\cite{lin2020software}.

\subsubsection{Dataset Preparation}

We used several vulnerability datasets to verify our model 
and compared the performance with baseline models. 

For \textbf{RQ1} to \textbf{RQ4}, we chose the 
\textit{Software Assurance Reference Dataset} (\textbf{SARD}), 
which is a widely used vulnerability database with 
a large set of synthetic programs~\cite{funded,deepwukong,lin2020software}. 
In \textbf{SARD}, a program is labelled as good (not vulnerable), 
bad (vulnerable) or mix (vulnerable with patched updates). 
For vulnerable programs, \textbf{SARD} describes the vulnerability and the vulnerability type 
in \textbf{CWE-ID} (\textbf{C}ommon \textbf{W}eakness \textbf{E}numeration \textbf{ID}entifier) formats. 
It also contains the human-crafted annotations in the program as supplementary information of the codes. 
We used a total of \textcolor{black}{22} categories for C/C++, Java and PHP, \textcolor{black}{of which are the 10 most common types in 2022 \textbf{CWE} Top 25 Most Dangerous Software Weaknesses\footnote{https://cwe.mitre.org/top25/archive/2022/2022\_cwe\_top25.html} and the rest 12 are most typical of other types.}
\textcolor{black}{Regarding the sample numbers of each type in \textbf{CWE} we present and explain in detail in Table \ref{tab:dataset}.}
\textcolor{black}{In each type, we selected 80\% of the sample size as the training dataset and 20\% as the test dataset.}
These categories are harvested from \textbf{SARD}, 
which is comprehensive and covers most of the vulnerability types.

\begin{figure}[htp!]
\vspace{-5mm}
\centering
\includegraphics[width=0.95\linewidth]{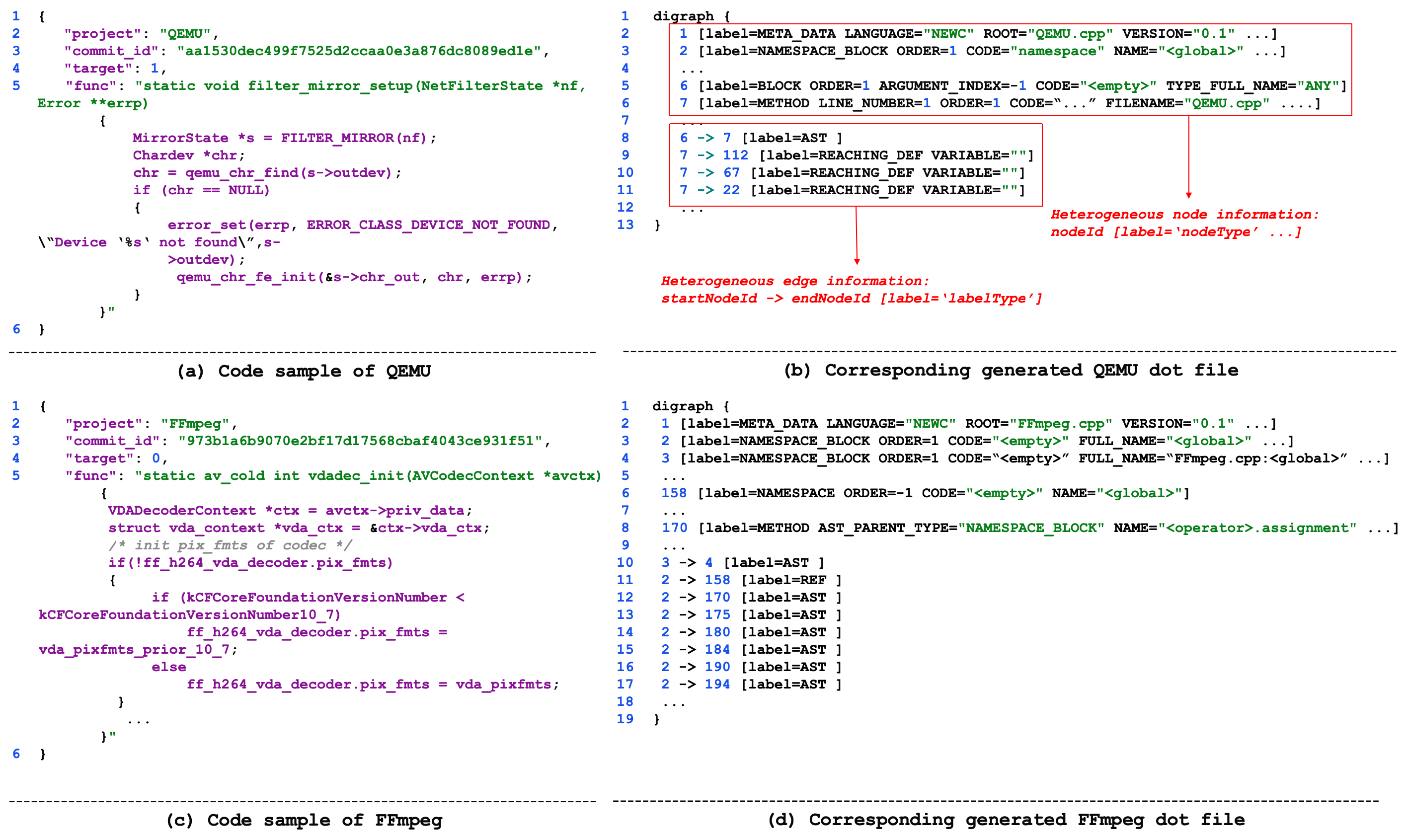}
\captionsetup{justification=centering}
\caption{\revise{Code sample of QEMU and FFmpeg, along with corresponding generated dot file.}}
\label{fig:data_pre_1}
\end{figure}

For \textbf{RQ5}, we leveraged three real-world open-source projects, 
\textbf{\textit{FFmpeg}}\footnote{https://ffmpeg.org/}, 
\textbf{\textit{QEMU}}\footnote{https://www.qemu.org/} 
and 
\textbf{\textit{Wireshark}}\footnote{https://samate.nist.gov/SARD/test-suites/8}.
These three large open-source projects are written in C, involving many contributions and code commits from software developers. 
The labels of \textbf{\textit{FFmpeg}}, \textbf{\textit{Wireshark}} and \textbf{\textit{QEMU}} 
are based on vulnerability-fix commits or non-vulnerability fix commits of these projects. 
The vulnerability fix commits (VFCs) are the code commits that fix a potential vulnerability of a function/method, while the non-vulnerability-fix commits (non-VFCs) are commits considered less relevant to fix vulnerabilities. The detailed statistics and descriptions of the datasets can be found in Table~\ref{realworldstatstics}.

\begin{figure}[htp!]
\centering
\includegraphics[width=0.95\linewidth]{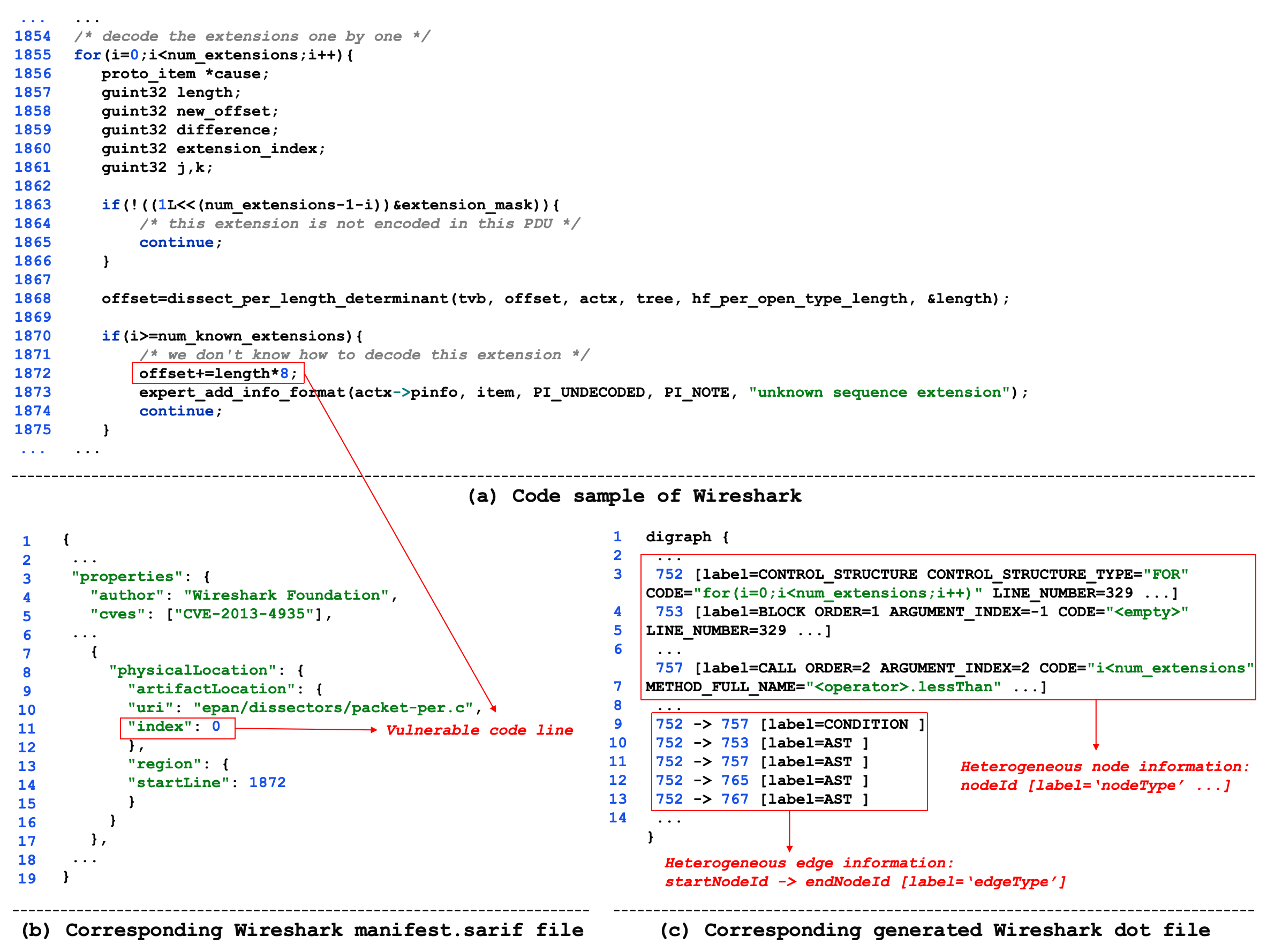}
\captionsetup{justification=centering}
\caption{\revise{Code sample of Wireshark and manifest.sarif file, along with the corresponding dot file.}}
\label{fig:data_pre_2}
\end{figure}

\revise{Code generation varies slightly across these real-world projects due to the distinctive presentation of code in each of them. For QEMU and FFmpeg, the code data are stored in the format of JSON, where the code snippets are under the “func” key and the labels are under the “target” key (shown in Figure~\ref{fig:data_pre_1}). Based on that, we developed a Python script to extract the code snippet as well as its labels iteratively, and use Joern to convert the respective code snippet into Code Property Graph (CPG) stored in the dot file. Taking Figure~\ref{fig:data_pre_1} (b) as an example, the dot file encompasses varying node and edge information, as well as its label indicating the abnormality, resulting in the heterogeneous graph that can be encoded for vulnerability detection tasks.  Regarding the Wireshark project, the codes are stored in a similar way to the SARD dataset, where each code file is accompanied by a corresponding manifest.sarif file, including some artifact information such as authors, index of the file, line of vulnerability, etc. (shown in Figure~\ref{fig:data_pre_2} (a) and (b)). The CPG of codes in Wireshark is generated in the same way as QEMU and FFmpeg using Joern.}

\begin{table*}[t!]
\vspace{-3mm}
\caption{Results of the comparison with different baselines on SARD (Accuracy and F1)} 
\label{tab:results_25}
\scalebox{0.85}{

\begin{tabular}{|l|ll|ll|ll|ll||ll|ll|}
\hline
\multirow{3}{*}{\diagbox[]{Dataset}{Metric}{Model}} & \multicolumn{2}{l|}{\multirow{2}{*}{DEVIGN}} & \multicolumn{2}{l|}{\multirow{2}{*}{LIN et al.}} & \multicolumn{2}{l|}{\multirow{2}{*}{FUNDED}}       & \multicolumn{2}{l||}{\multirow{2}{*}{DeepWukong}} & \multicolumn{2}{l|}{\multirow{2}{*}{DSHGTnoAnno}}  & \multicolumn{2}{l|}{\multirow{2}{*}{DSHGT}}        \\
                        & \multicolumn{2}{l|}{}                        & \multicolumn{2}{l|}{}                            & \multicolumn{2}{l|}{}                              & \multicolumn{2}{l||}{}                            & \multicolumn{2}{l|}{}                              & \multicolumn{2}{l|}{}                              \\ \cline{2-13} 
                        & \multicolumn{1}{l|}{ACC}             & F1    & \multicolumn{1}{l|}{ACC}               & F1      & \multicolumn{1}{l|}{ACC}           & F1            & \multicolumn{1}{l|}{ACC}               & F1      & \multicolumn{1}{l|}{ACC}           & F1            & \multicolumn{1}{l|}{ACC}           & F1            \\ \hline
CWE-119                 & \multicolumn{1}{l|}{0.79}            & 0.81  & \multicolumn{1}{l|}{\textbf{0.85}}     & 0.78    & \multicolumn{1}{l|}{0.83}          & 0.85          & \multicolumn{1}{l|}{0.82}              & 0.84    & \multicolumn{1}{l|}{0.80}          & 0.86          & \multicolumn{1}{l|}{0.84}          & \textbf{0.87} \\ \hline
CWE-400                 & \multicolumn{1}{l|}{\textbf{0.84}}   & 0.81  & \multicolumn{1}{l|}{0.79}              & 0.76    & \multicolumn{1}{l|}{\textbf{0.84}} & 0.80          & \multicolumn{1}{l|}{\textbf{0.84}}     & 0.79    & \multicolumn{1}{l|}{0.82}          & 0.83          & \multicolumn{1}{l|}{0.83}          & \textbf{0.85} \\ \hline
CWE-404                 & \multicolumn{1}{l|}{0.84}            & 0.82  & \multicolumn{1}{l|}{0.83}              & 0.74    & \multicolumn{1}{l|}{0.81}          & 0.85          & \multicolumn{1}{l|}{0.83}              & 0.84    & \multicolumn{1}{l|}{0.83}          & 0.88          & \multicolumn{1}{l|}{\textbf{0.88}} & \textbf{0.90} \\ \hline
CWE-369                 & \multicolumn{1}{l|}{0.83}            & 0.78  & \multicolumn{1}{l|}{0.82}              & 0.80    & \multicolumn{1}{l|}{0.86}          & 0.84          & \multicolumn{1}{l|}{\textbf{0.91}}     & 0.87    & \multicolumn{1}{l|}{0.90}          & \textbf{0.91} & \multicolumn{1}{l|}{0.89}          & 0.88          \\ \hline
CWE-191                 & \multicolumn{1}{l|}{0.82}            & 0.76  & \multicolumn{1}{l|}{0.80}              & 0.73    & \multicolumn{1}{l|}{\textbf{0.87}} & 0.90          & \multicolumn{1}{l|}{0.75}              & 0.81    & \multicolumn{1}{l|}{0.81}          & 0.87          & \multicolumn{1}{l|}{0.85}          & \textbf{0.91} \\ \hline
CWE-476                 & \multicolumn{1}{l|}{0.91}            & 0.87  & \multicolumn{1}{l|}{0.89}              & 0.83    & \multicolumn{1}{l|}{0.83}          & 0.87          & \multicolumn{1}{l|}{\textbf{0.90}}     & 0.86    & \multicolumn{1}{l|}{0.85}          & 0.84          & \multicolumn{1}{l|}{0.86}          & \textbf{0.89} \\ \hline
CWE-467                 & \multicolumn{1}{l|}{0.79}            & 0.84  & \multicolumn{1}{l|}{0.87}              & 0.81    & \multicolumn{1}{l|}{0.85}          & 0.86          & \multicolumn{1}{l|}{0.88}              & 0.86    & \multicolumn{1}{l|}{0.86}          & 0.83          & \multicolumn{1}{l|}{\textbf{0.90}} & \textbf{0.87} \\ \hline
CWE-78                  & \multicolumn{1}{l|}{0.82}            & 0.84  & \multicolumn{1}{l|}{\textbf{0.89}}     & 0.79    & \multicolumn{1}{l|}{0.83}          & \textbf{0.86} & \multicolumn{1}{l|}{0.84}              & 0.83    & \multicolumn{1}{l|}{0.84}          & 0.86          & \multicolumn{1}{l|}{0.85}          & 0.84          \\ \hline
CWE-772                 & \multicolumn{1}{l|}{0.83}            & 0.77  & \multicolumn{1}{l|}{0.85}              & 0.81    & \multicolumn{1}{l|}{0.86}          & 0.87          & \multicolumn{1}{l|}{0.86}              & 0.83    & \multicolumn{1}{l|}{0.86}          & 0.88          & \multicolumn{1}{l|}{\textbf{0.90}} & \textbf{0.88} \\ \hline
CWE-190                 & \multicolumn{1}{l|}{0.86}            & 0.83  & \multicolumn{1}{l|}{0.83}              & 0.79    & \multicolumn{1}{l|}{0.86}          & 0.84          & \multicolumn{1}{l|}{0.87}              & 0.83    & \multicolumn{1}{l|}{0.85}          & 0.82          & \multicolumn{1}{l|}{\textbf{0.92}} & \textbf{0.87} \\ \hline
CWE-770                 & \multicolumn{1}{l|}{0.87}            & 0.84  & \multicolumn{1}{l|}{0.89}              & 0.80    & \multicolumn{1}{l|}{0.85}          & 0.87          & \multicolumn{1}{l|}{0.86}              & 0.87    & \multicolumn{1}{l|}{0.85}          & 0.86          & \multicolumn{1}{l|}{\textbf{0.90}} & \textbf{0.89} \\ \hline
CWE-666                 & \multicolumn{1}{l|}{0.85}            & 0.84  & \multicolumn{1}{l|}{0.88}              & 0.86    & \multicolumn{1}{l|}{0.89}          & 0.90          & \multicolumn{1}{l|}{0.87}              & 0.92    & \multicolumn{1}{l|}{0.86}          & 0.91          & \multicolumn{1}{l|}{\textbf{0.90}} & \textbf{0.93} \\ \hline
CWE-665                 & \multicolumn{1}{l|}{0.83}            & 0.87  & \multicolumn{1}{l|}{0.90}              & 0.79    & \multicolumn{1}{l|}{0.93}          & 0.88          & \multicolumn{1}{l|}{0.92}              & 0.89    & \multicolumn{1}{l|}{0.92}          & 0.92          & \multicolumn{1}{l|}{\textbf{0.94}} & \textbf{0.92} \\ \hline
CWE-758                 & \multicolumn{1}{l|}{0.84}            & 0.87  & \multicolumn{1}{l|}{0.86}              & 0.83    & \multicolumn{1}{l|}{0.84}          & 0.88          & \multicolumn{1}{l|}{0.87}              & 0.92    & \multicolumn{1}{l|}{0.85}          & 0.89          & \multicolumn{1}{l|}{\textbf{0.91}} & \textbf{0.93} \\ \hline
CWE-469                 & \multicolumn{1}{l|}{0.75}            & 0.79  & \multicolumn{1}{l|}{0.78}              & 0.76    & \multicolumn{1}{l|}{\textbf{0.86}} & 0.83          & \multicolumn{1}{l|}{0.76}              & 0.79    & \multicolumn{1}{l|}{\textbf{0.83}} & 0.84          & \multicolumn{1}{l|}{0.83}          & \textbf{0.86} \\ \hline
CWE-676                 & \multicolumn{1}{l|}{0.84}            & 0.80  & \multicolumn{1}{l|}{0.84}              & 0.75    & \multicolumn{1}{l|}{\textbf{0.92}} & \textbf{0.91} & \multicolumn{1}{l|}{0.89}              & 0.83    & \multicolumn{1}{l|}{0.86}          & 0.89          & \multicolumn{1}{l|}{0.90}          & \textbf{0.91} \\ \hline
CWE-834                 & \multicolumn{1}{l|}{0.70}            & 0.62  & \multicolumn{1}{l|}{0.76}              & 0.74    & \multicolumn{1}{l|}{0.84}          & 0.79          & \multicolumn{1}{l|}{0.74}              & 0.72    & \multicolumn{1}{l|}{0.83}          & 0.76          & \multicolumn{1}{l|}{\textbf{0.87}} & \textbf{0.82} \\ \hline
CWE-79                  & \multicolumn{1}{l|}{0.82}            & 0.84  & \multicolumn{1}{l|}{0.84}              & 0.85    & \multicolumn{1}{l|}{0.85}          & 0.83          & \multicolumn{1}{l|}{0.86}              & 0.88    & \multicolumn{1}{l|}{0.86}          & 0.87          & \multicolumn{1}{l|}{\textbf{0.88}} & \textbf{0.90} \\ \hline
CWE-89                  & \multicolumn{1}{l|}{0.85}            & 0.82  & \multicolumn{1}{l|}{0.76}              & 0.80    & \multicolumn{1}{l|}{0.83}          & 0.85          & \multicolumn{1}{l|}{0.81}              & 0.84    & \multicolumn{1}{l|}{\textbf{0.87}} & 0.85          & \multicolumn{1}{l|}{\textbf{0.87}} & \textbf{0.87} \\ \hline
CWE-416                 & \multicolumn{1}{l|}{0.83}            & 0.85  & \multicolumn{1}{l|}{0.80}              & 0.81    & \multicolumn{1}{l|}{0.84}          & 0.87          & \multicolumn{1}{l|}{0.79}              & 0.84    & \multicolumn{1}{l|}{0.85}          & 0.88          & \multicolumn{1}{l|}{\textbf{0.88}} & \textbf{0.90} \\ \hline
CWE-20                  & \multicolumn{1}{l|}{0.84}            & 0.88  & \multicolumn{1}{l|}{0.80}              & 0.83    & \multicolumn{1}{l|}{0.87}          & 0.89          & \multicolumn{1}{l|}{0.86}              & 0.87    & \multicolumn{1}{l|}{0.89}          & 0.90          & \multicolumn{1}{l|}{\textbf{0.90}} & \textbf{0.92} \\ \hline
CWE-125                 & \multicolumn{1}{l|}{0.81}            & 0.85  & \multicolumn{1}{l|}{0.78}              & 0.83    & \multicolumn{1}{l|}{0.82}          & 0.87          & \multicolumn{1}{l|}{0.84}              & 0.84    & \multicolumn{1}{l|}{0.85}          & 0.86          & \multicolumn{1}{l|}{\textbf{0.86}} & \textbf{0.89} \\ \hline
\end{tabular}

}
\vspace{-3mm}
\end{table*}

\subsection{Performance Analysis on \textbf{SARD} (RQ1)}
We first verified the effectiveness of the proposed model and other baselines on a number of CWE vulnerability types, 
from which the models are trained and tested on \textbf{SARD} synthetic code samples. 
Note that the training and testing data ratio is 80\% and 20\%, respectively. 
\revised{Table~\ref{tab:results_25} reports the overall performance of each model on different vulnerability types using Accuracy and F1. The other metrics on Precision and Recall can be found in Table~\ref{otherTwo} in the appendix.
In general, even without incorporating the semantic meaning of code comments into the model (\tool$_{noAnno}$), 
our proposed model achieves promising results on almost all vulnerability types. 
\tool$_{noAnno}$ is the variant of \tool~only training on code representation graph using vulnerability oracle. }
Devign, Lin et al., DeepWukong, and \tool$_{noAnno}$ give low F1 and accuracy score on the CWE-834 dataset, 
which describes an ``\textit{Excessive Iteration}'' vulnerability that 
leads to the over-consumption of resources and can be adversely affected by attackers. 
This type of vulnerability presents no clear sign of vulnerable code patterns 
and is hardly identified by learning solely on code graph representations or code tokens. 
Note that the method in DeepWuKong of constructing Extracted Flow Graphs (XFG) is based on prior knowledge of vulnerable code locations, but this is infeasible to retrieve that in real-world scenarios. In order to facilitate comparisons on an equal footing, we created a dataset for each method individually, which could lead to slightly different results compared to those presented in the original paper of DeepWukong.
\revised{\tool, on the other hand, also leverages the semantic code comment information of the programs to enhance the robustness of the detection ability, 
thus achieving much better results compared with \tool$_{noAnno}$ and other baselines.} 
It can also be observed that both Devign and Lin et al. perform undesirable on CWE-469, 
in which the vulnerability is caused by the misuse of the pointer variable, while \tool~is the best-performing model on this dataset, 
indicating that the intra-program dependence in the CPG provides sufficient information when modelling the code graph in this case. 
FUNDED achieves marginal performance gain compared with \tool~and DeepWukong on CWE-676. 
We discover that this dataset relates to ``Use of Potentially Dangerous Function'' vulnerability, 
which can be identified by models capable of encoding control flow information. 
Therefore, FUNDED, DeepWukong and \tool~achieve similar results on this dataset. 
Lin et al. achieves the best Accuracy score on CWE-78, indicating high True Positive and Negative numbers. 
Yet it reports a low F1 score with poor results on Precision and Recall.
Overall, \tool~achieves the best results on most of the tested datasets, 
with an average 88.05\% on accuracy and 88.35\% on F1.

\begin{table*}[t!]
\caption{\textcolor{black}{Results of T-test}} 
\label{tab:T-test}
\scalebox{0.7}{

\begin{tabular}{|c|c|c|c|c|c|c|}
\hline
\diagbox[]{Comparison}{T-test data}                                &                      & Mean   & Standard deviation & T-value                 & Null Hypothesis(H0)  & Alternative Hypothesis(H1) \\ \hline
\multirow{4}{*}{\tool-DEVIGN}      & \multirow{2}{*}{ACC} & 0.8786 & 0.0329             & \multirow{2}{*}{2.0262} & \multirow{2}{*}{\XSolidBrush} & \multirow{2}{*}{\Checkmark}       \\ \cline{3-4}
                                   &                      & 0.8205 & 0.0489             &                         &                      &                            \\ \cline{2-7} 
                                   & \multirow{2}{*}{F1}  & 0.8791 & 0.0333             & \multirow{2}{*}{2.8975} & \multirow{2}{*}{\XSolidBrush} & \multirow{2}{*}{\Checkmark}       \\ \cline{3-4}
                                   &                      & 0.7909 & 0.0385             &                         &                      &                            \\ \hline
\multirow{4}{*}{\tool-LIN et al.}  & \multirow{2}{*}{ACC} & 0.8795 & 0.0337             & \multirow{2}{*}{2.2596} & \multirow{2}{*}{\XSolidBrush} & \multirow{2}{*}{\Checkmark}       \\ \cline{3-4}
                                   &                      & 0.8518 & 0.0274             &                         &                      &                            \\ \cline{2-7} 
                                   & \multirow{2}{*}{F1}  & 0.8845 & 0.0302             & \multirow{2}{*}{2.4167} & \multirow{2}{*}{\XSolidBrush} & \multirow{2}{*}{\Checkmark}       \\ \cline{3-4}
                                   &                      & 0.8586 & 0.0322             &                         &                      &                            \\ \hline
\multirow{4}{*}{\tool-FUNDED}      & \multirow{2}{*}{ACC} & 0.8827 & 0.0361             & \multirow{2}{*}{2.561}  & \multirow{2}{*}{\XSolidBrush} & \multirow{2}{*}{\Checkmark}       \\ \cline{3-4}
                                   &                      & 0.8331 & 0.0546             &                         &                      &                            \\ \cline{2-7} 
                                   & \multirow{2}{*}{F1}  & 0.8841 & 0.0367             & \multirow{2}{*}{1.520}  & \multirow{2}{*}{\Checkmark} & \multirow{2}{*}{\XSolidBrush}       \\ \cline{3-4}
                                   &                      & 0.8573 & 0.0388             &                         &                      &                            \\ \hline
\multirow{4}{*}{\tool-DeepWukong}  & \multirow{2}{*}{ACC} & 0.8645 & 0.0184             & \multirow{2}{*}{2.6197} & \multirow{2}{*}{\XSolidBrush} & \multirow{2}{*}{\Checkmark}       \\ \cline{3-4}
                                   &                      & 0.8486 & 0.0281             &                         &                      &                            \\ \cline{2-7} 
                                   & \multirow{2}{*}{F1}  & 0.8841 & 0.0367             & \multirow{2}{*}{1.520}  & \multirow{2}{*}{\Checkmark} & \multirow{2}{*}{\XSolidBrush}       \\ \cline{3-4}
                                   &                      & 0.8361 & 0.0517             &                         &                      &                            \\ \hline
\multirow{4}{*}{\tool-DSHGTnoAnno} & \multirow{2}{*}{ACC} & 0.8682 & 0.018              & \multirow{2}{*}{2.6006} & \multirow{2}{*}{\XSolidBrush} & \multirow{2}{*}{\Checkmark}       \\ \cline{3-4}
                                   &                      & 0.8536 & 0.0197             &                         &                      &                            \\ \cline{2-7} 
                                   & \multirow{2}{*}{F1}  & 0.8800 & 0.012              & \multirow{2}{*}{1.3149} & \multirow{2}{*}{\Checkmark} & \multirow{2}{*}{\XSolidBrush}       \\ \cline{3-4}
                                   &                      & 0.8636 & 0.0194             &                         &                      &                            \\ \hline
\end{tabular}

}
\end{table*}

\textcolor{black}{
In order to more clearly and accurately demonstrate the advantages of the \tool, we also performed a quantitative analysis of the experiment data (Table \ref{tab:results_25}), using independent samples t-test.
We tested \tool~against DEVIGN, LIN et al., FUNDED, DeepWukong, and \tool$_{noAnno}$ separately.
In each test the comparisons were made separately for Accuracy and F1. The quantitative analysis is used to show whether \tool~is due to the comparison schemes.
We use $\tool.ACC$ to denote a set of experiment results.
Use $\Bar{X}_{\tool.ACC}$ to denote the mean of $\tool.ACC$, $S_{\tool.ACC}$ to denote the standard deviation, and $N_{\tool.ACC}$ to denote the number of samples:
}
\begin{equation}
\label{MeanDeviationNum}
\begin{split}
    & \textcolor{black}{\Bar{X}_{\tool.ACC}=\frac{1}{N_{\tool.ACC}}\sum^{N_{\tool.ACC}}_{i=1}\Bar{X}^{i}_{\tool.ACC}} \\
    & \textcolor{black}{S_{\tool.ACC}=\sqrt{\frac{1}{N_{\tool.ACC}-1}\sum^{N_{\tool.ACC}}_{i=1}(\Bar{X}^{i}_{\tool.ACC}-\Bar{X}_{\tool.ACC})^{2}}}
\end{split}
\end{equation}

\begin{figure}[t!]
\centering
\includegraphics[width=0.78\linewidth]{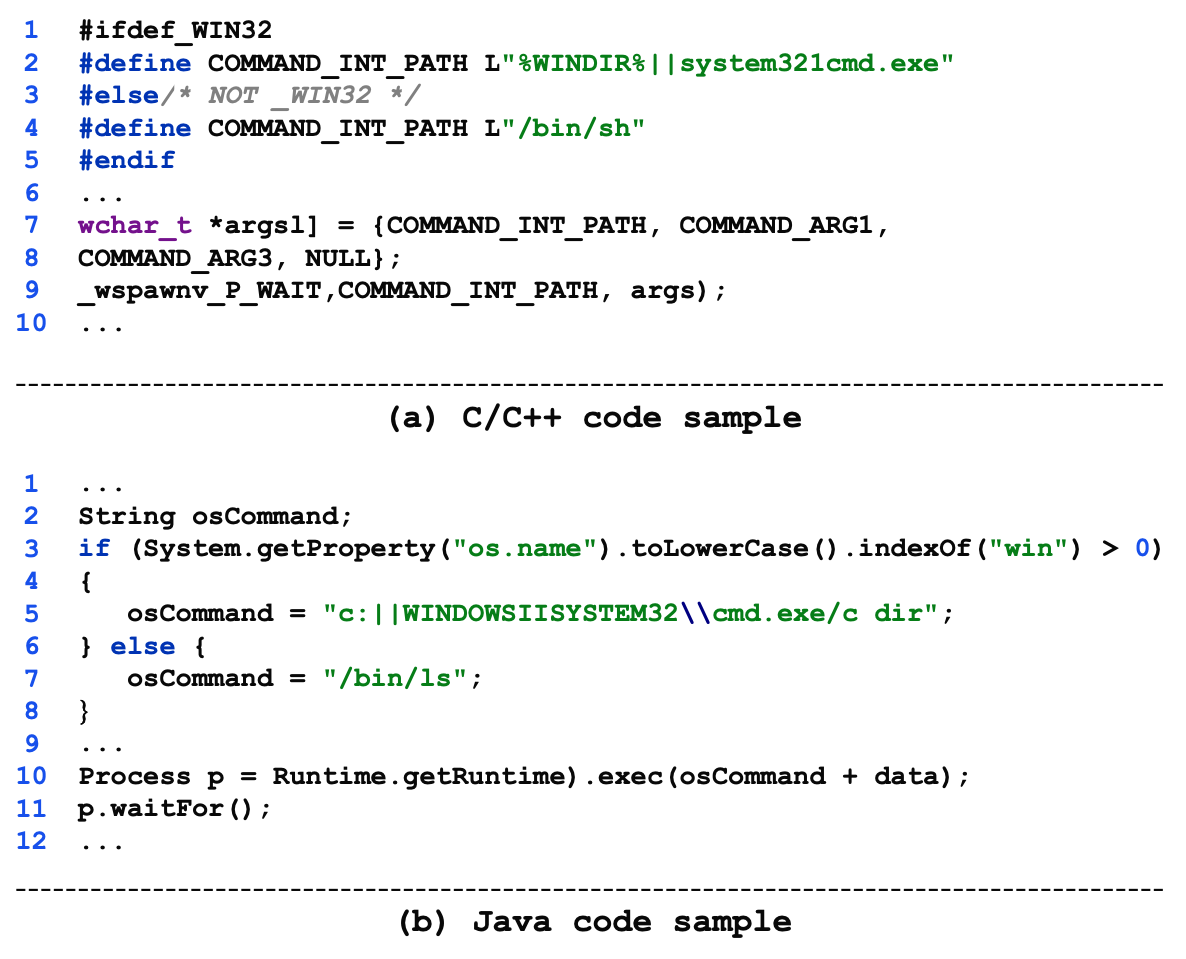}
\captionsetup{justification=centering}
\caption{Both C/C++ and Java code samples with ``OS Command Injection'' vulnerability on CWE-78}
\label{fig:codesample}
\end{figure}

When we compare $DEVIGN.ACC$ and $\tool.ACC$, we can construct from the definition of the t-distribution:
\begin{equation}
\label{MeanDeviationNum}
    \textcolor{black}{\frac{\Bar{X}_{\tool.ACC}-\Bar{X}_{DEVIGN.ACC}}{\sqrt{\frac{S^{2}_{\tool.ACC}}{N_{\tool.ACC}}+\frac{S^{2}_{DEVIGN.ACC}}{N_{DEVIGN.ACC}}}}\sim t(V)} 
\end{equation}
where the degrees of freedom $V$ of this t-distribution are related to the degrees of freedom $V_{\tool.ACC}$  and $V_{DEVIGN.ACC}$:

\begin{equation}
\label{MeanDeviationNum}
    \textcolor{black}{V\approx\frac{(\frac{S^{2}_{\tool.ACC}}{N_{\tool.ACC}}+\frac{S^{2}_{DEVIGN.ACC}}{N_{DEVIGN.ACC}})^{2}}{\frac{S^{4}_{\tool.ACC}}{N^{2}_{\tool.ACC}\cdot V_{\tool.ACC}}+\frac{S^{4}_{DEVIGN.ACC}}{N^{2}_{DEVIGN.ACC}\cdot V_{DEVIGN.ACC}}}} 
\end{equation}

\textcolor{black}{We performed a quantitative analysis of the experimental results based on the aforementioned $\Bar{X}$, $S$, $V$ for the T-test, and the results are presented in Table \ref{tab:T-test}.}
\textcolor{black}{We first set a common significance level of 0.05 and degrees of freedom as 42.}
\textcolor{black}{Then we formulated two hypotheses and set a common significance level of 0.05:}
\begin{itemize}
    \item \textcolor{black}{Null Hypothesis (H0): There is no significant difference between the means of the two samples, i.e., $\mu1=\mu2$.}
    \item \textcolor{black}{Alternative Hypothesis (H1): There is a significant difference between the means of the two samples, i.e., $\mu1\neq \mu2$.}
\end{itemize}

It can be observed that we can reject the Null Hypothesis and accept the Alternative Hypothesis except for F1 in comparison to \tool~and FUNDED, F1 in comparison to \tool~and DeepWukong and F1 in comparison to \tool~and \textit{DSHGTnoAnno}.
In all three of the above comparisons, we cannot quantitatively determine that \tool~is better at a significance level of 0.05. Nevertheless, upon a direct and close examination of the results, it becomes evident that \tool~ outperforms the comparison programs in all three metrics.

While the enhancement of the proposed method is modest in comparison to the baselines, our primary objective is to enhance the generalizability of our detection model, as evidenced by the subsequent experiments.

\begin{figure}[t!]
\centering
\includegraphics[width=0.99\linewidth]{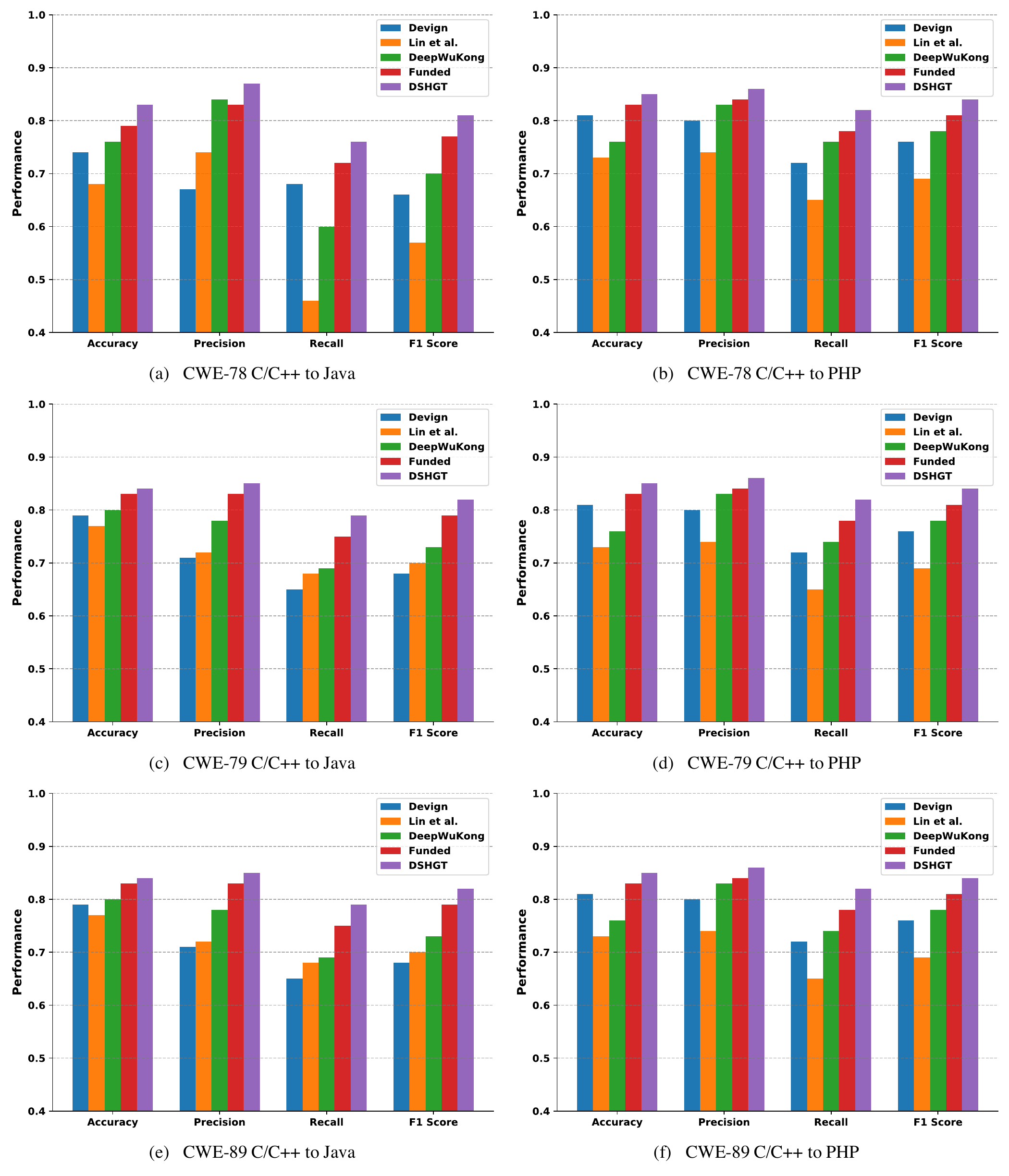}
\captionsetup{justification=centering}
\caption{\textcolor{black}{Knowledge transferring capability from C/C++ to other programming languages}}
\label{fig:transfer}
\vspace{-5mm}
\end{figure}

\subsection{Transferability on Other Programming Languages (RQ2)}
We experimented with the transferring capability of our proposed model against other baselines. 
To achieve that, we adopted the transfer learning technique by keeping the model structure and weights trained on C/C++ dataset, 
and fine-tuned that on other programming languages for vulnerability detection tasks. 
It is proven feasible in~\cite{funded} that the model trained in one programming language domain 
can preserve the knowledge of vulnerable code patterns and thus be used to detect similar vulnerabilities in other programming languages.

In this case, we first trained our model and other baselines on C/C++ datasets. To transfer the knowledge learned from C/C++, we froze the architecture of HGT in our framework and only fine-tuned the final MLP classifier (2 linear layers and 1 output layer) for 10 epochs. Note that we first transformed different programming languages into Control Property Graphs (CPG) in a languages-agnostic manner and then employed \textit{DSHGT} on these graphs so that \textit{DSHGT} can be easily employed to detect similar vulnerabilities in other programming languages. We selected three types of vulnerabilities, CWE-78, CWE-79, and CWE-89, to further verify the transferability of our approach, as these are the only vulnerability types that co-occur across the C/C++, Java, and PHP programming languages in the dataset. We considered two cases that use transfer learning, C/C++ to Java and C/C++ to PHP. 
Figure~\ref{fig:transfer} shows the result of the models' transferability \textcolor{black}{on CWE-78, CWE-79 and CWE-89}. It can be observed that our pre-trained model can better capture the prior vulnerable code patterns 
and achieve promising results when applied to other programming languages, 
with 84\% accuracy from C to Java and 88\% accuracy from C to PHP.

\begin{table}[t!]
\captionsetup{justification=centering}
\caption{Analysis on HGT layer \\and training time cost on CWE-119}
\label{tab:hgtlayer}
\scalebox{0.99}{
\begin{tabular}{|l|lllll|}
\hline
\multirow{2}{*}{Metric} & \multicolumn{5}{c|}{HGT Layer Depths}                                                                                   \\ \cline{2-6} 
                        & \multicolumn{1}{l|}{1}    & \multicolumn{1}{l|}{2}    & \multicolumn{1}{l|}{3}     & \multicolumn{1}{l|}{4}     & 5     \\ \hline
F1                      & \multicolumn{1}{l|}{0.79} & \multicolumn{1}{l|}{0.87} & \multicolumn{1}{l|}{\textbf{0.89}}  & \multicolumn{1}{l|}{0.88}  & 0.85  \\ \hline
Training Time Cost/h    & \multicolumn{1}{l|}{4.83} & \multicolumn{1}{l|}{9.45} & \multicolumn{1}{l|}{14.88} & \multicolumn{1}{l|}{19.81} & 23.64 \\ \hline
\end{tabular}
}
\end{table}

\begin{figure}[t!]
\centering
\includegraphics[width=0.8\linewidth]{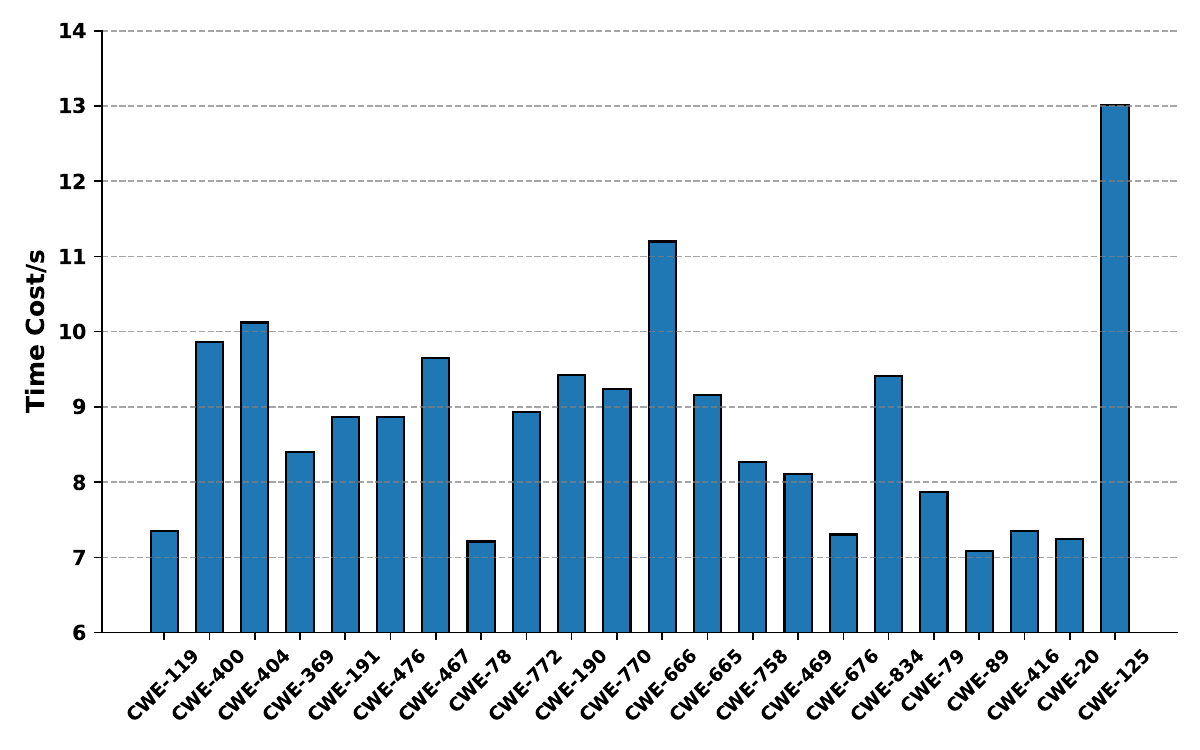}
\captionsetup{justification=centering}
\caption{CPG building time}
\label{fig:cpgTime}
\end{figure}

We further explored the rationale behind the effectiveness of using transfer learning. 
It can be observed from Figure~\ref{fig:codesample} that both C/C++ and Java code samples 
in CWE-78 construct an OS command (highlighted in red) using externally-influenced input 
from an upstream component without validating the special element that could harm the system, 
thus are under threat of command inject attacks. 
In the CPG of both code samples, a control flow edge should exist if this command variable is validated before being used by other threads, 
thus presenting a similar code pattern regardless of language syntax. 
The result verifies that our model can better capture both contextual semantics 
and underlying structure (syntax, control- and data-flow) of the codes with the help of CPG and the corresponding heterogeneous graph learning, 
thus able to preserve the language-agnostic vulnerable code patterns.

\begin{figure}[t!]
\centering
\includegraphics[width=0.9\linewidth]{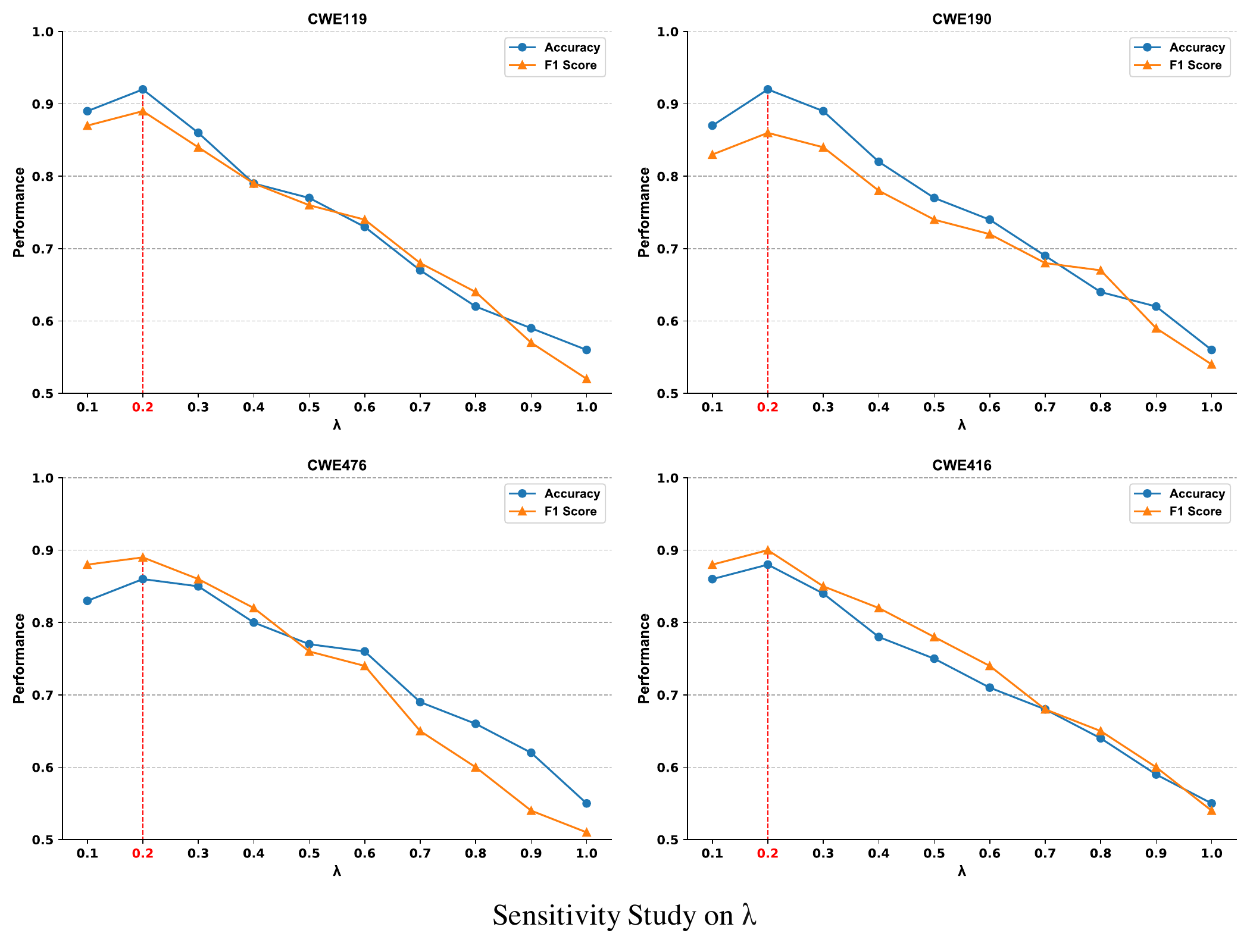}
\caption{Change of Acc and F1 along with $\lambda$ on CWE-119, CWE-190, CWE-476 and CWE-416}
\label{fig:sensitivity}
\end{figure}

\subsection{Impact of Dual-supervisors (RQ3)}
\label{sec:rq3}
To answer RQ3, we studied the necessity of designing multi-task learning with dual supervisors 
to enhance the performance of \tool~as well as how to balance between two supervisors responsible 
for dealing with the vulnerability label oracle and the \textcolor{black}{code comment} oracle, 
in which the textual \textcolor{black}{code comments} are used as supplementary information. 
Note the use of auxiliary annotations of code snippets is proven helpful for enhancing comprehension of code semantics and has already been used in tasks like code completion~\cite{annotation1}, 
code summarization~\cite{annotation2}, and code retrival~\cite{annotation3}. For example, PLBART~\cite{ahmad-etal-2021-unified} demonstrates promising performance in code summarization and generation by pretraining a sequence-to-sequence network on code functions and their relevant natural language descriptions. Similarly, CodeBERT~\cite{feng2020codebert} and GraphCodeBERT~\cite{guo2021graphcodebert} implement transformer-based neural architectures on corpora composed of code functions and their respective code comments, with the aim of deriving high-quality embedding representations of code. We thus hypothesize that the graph code embedding should contain rich semantic information capable of summarizing the code snippets in code comment formats. 
\revised{Apart from that, the sensitivity hyperparameter $\lambda$ was explored in the experiment to investigate the level of impact caused by contextual code comments when training the model, in which 0 indicates only training the model based on vulnerability labels, while 1 means only optimizing the model towards generating correct code comment summaries of codes (reflected in Eq.13). }

We first experimented on the HGT layer depth and compare the change of layer depth against detection performance and mode training cost. 
It is shown in Table~\ref{tab:hgtlayer} that the time cost increases with the increase of HGT layer depths, and F1 reaches the highest when layer depth is 3. 
We thus chose 3 as the layer depth setup for HGT. 
We also recorded CPG build times for each CWE type to study the time consumption of graph construction (as shown in Figure~\ref{fig:cpgTime}). It is observed clearly that the time consumption of graph construction, ranging from 7 to 13 seconds depending on number of code lines and the complexity of invocation among functions, is relatively small compared to the overall training period of \textit{DSHGT} (in the magnitude of hours).

\begin{figure}[t!]
\centering
\includegraphics[width=1\linewidth]{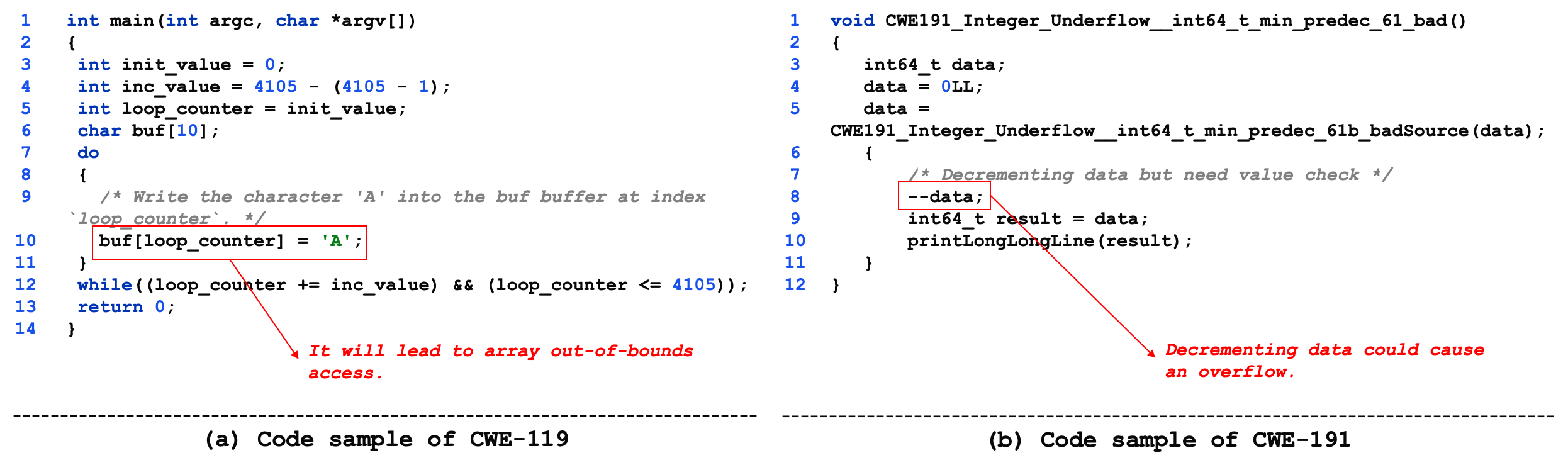}
\caption{\revise{Vulnerable code samples detected by DSHGT while Vulnerable code samples detected by DSHGT while overlooked by others}}
\label{fig:code_comment}
\end{figure}
Regarding the sensitivity of $\lambda$, we conducted experiments on four representative vulnerability types from the top-25 reported vulnerabilities in the year of 2023\footnote{https://cwe.mitre.org/top25/archive/2023/2023\_top25\_list.html}, including CWE-119 (Improper Restriction of Operations within the Bounds of a Memory Buffer), CWE-190 (Integer Overflow or Wraparound), CWE-476 (NULL Pointer Dereference), CWE-416 (Use After Free). \revised{Figure~\ref{fig:sensitivity} reveals the change of Accuracy and F1 scores along with $\lambda$.
Both scores on four datasets reach the highest when $\lambda$ = 0.2 and decrease afterwards on, 
indicating that over-reliance on code comments aggravates the vulnerability detection ability 
subject to the quality and number of code comments in the programs, while using semantic information in code comments is beneficial to some extent.} \revise{We provide two concrete examples to show the effectiveness of harnessing the auxiliary code comments. Figure~\ref{fig:code_comment} (a) demonstrates a potential buffer overflow vulnerability, where the character `A' is attempted to be written into the array buf with non-existent index 4105 (line 10 of the code). This simple code structure leads to an over-simplified code graph representation, thus is easy to be misclassified by simply using graph learning-based models. However, the code comment ``\textit{Writing outside the bounds of the `buf' array}” provides specific descriptions of this vulnerability in the code snippets, the semantics of which are then comprehended into the graph-level embedding through the training process and become helpful for determining whether the method is vulnerable. Another code example Figure~\ref{fig:code_comment} (b) illustrates the vulnerability caused by potential variable underflow. In this code sample, the variable data is initially declared as an int64\_t on line 3. Its value is then assigned by the function CWE191\_Integer\_Underflow\_\_int64\_t\_min\_predec\_61b\_badSource(), decrementing the variable data without a value check can lead to a potential underflow error owing to the possibility that the value of the variable data is set to the lower bound of the type int64\_t (line 8 of the code). The generated code graph fails to capture this information because  CWE191\_Integer\_Underflow\_\_int64\_t\_min\_predec\_61b\_badSource() lacks a return statement, resulting in the miss of both data and control edges in this case. In contrast, the dual-supervisor structure in DSHGT, which harnesses the code comments ``\textit{Decrementing data but needs value check}” in line 7, successfully comprehends natural descriptions of the functionality in this code snippet and consequently leads to correct prediction results. 
}

\begin{figure}[t!]
\centering
\includegraphics[width=0.8\linewidth]{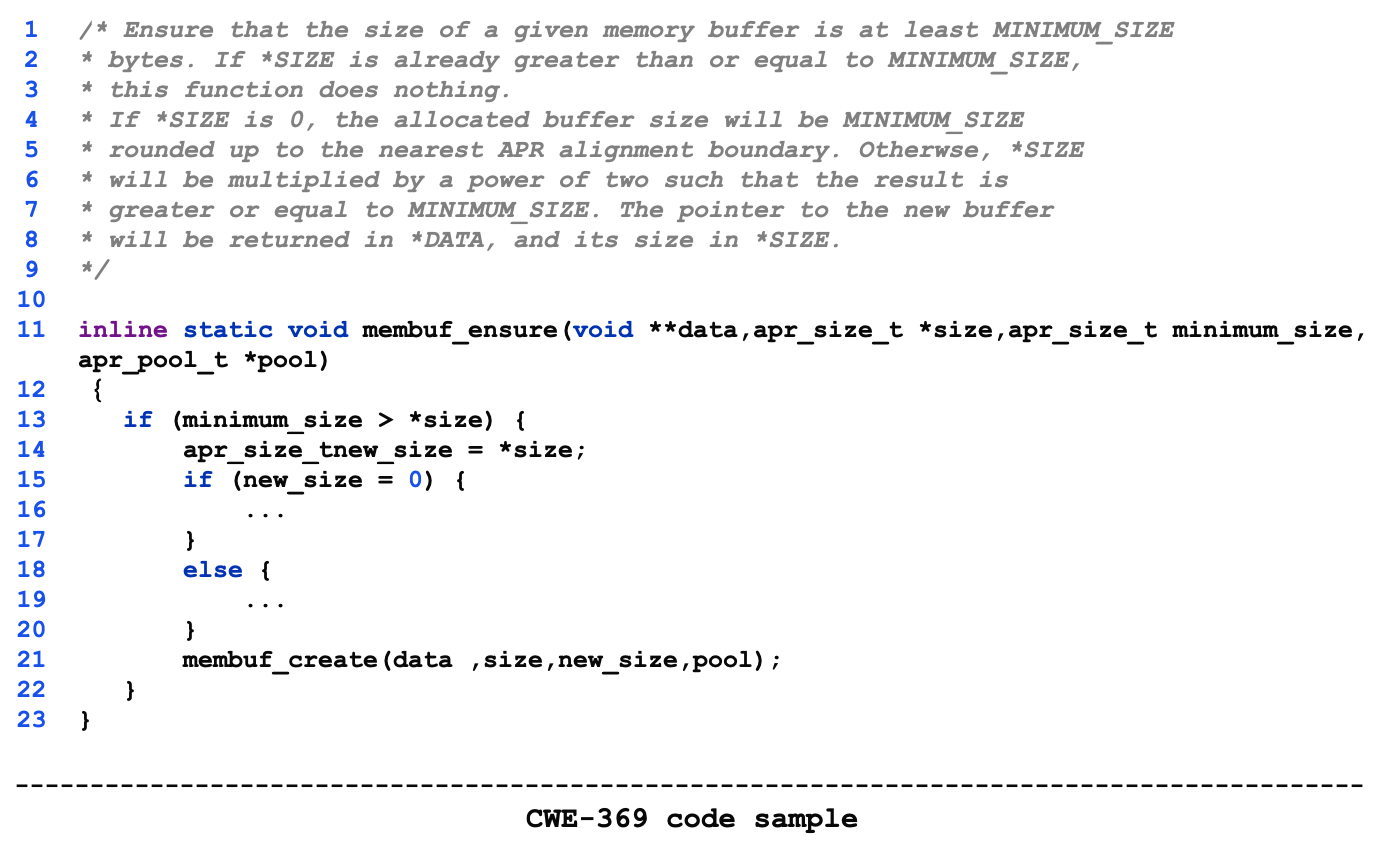}
\caption{CWE-369 code sample with redundant and noisy auxiliary comments}
\label{fig:code_sample_anno}
\end{figure}

\revised{However, it is not uncommon to have a mixture of good and bad code comments in the program. Taking CWE-119 as an example, we discover that some code comments 
like ``\textit{fixes the problem of not freeing the data in the source}'' have specific meanings or descriptions for the method, 
the semantics of which are then comprehended into the graph-level embedding through 
the training process and become helpful for determining whether the method is vulnerable, 
while there exists some code comments like ``\textit{use goodsource 
and badsink by changing the first GLOBAL\_CONST\_TRUE to GLOBAL\_CONST\_FALSE}'', which are not helpful. Another example can be found in CWE-369 (shown in Figure~\ref{fig:code_sample_anno}), where the code samples present redundant and noisy auxiliary comments, and the \textit{DSHGTnoAnno} thus performs better than \textit{DSHGT} in this case (also reflected in Table~\ref{tab:results_25}). Over-reliance on such code comments thus (large $\lambda$) deteriorates the vulnerability detection performance.}

\begin{figure}[t!]
\centering
\includegraphics[width=0.7\linewidth]{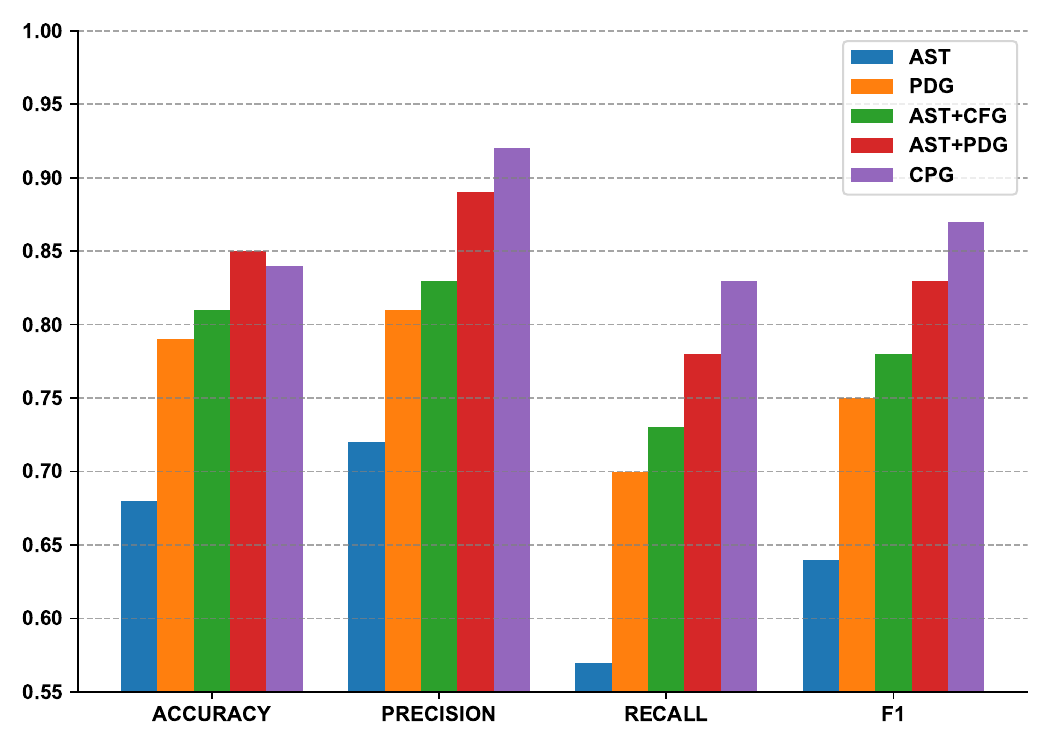}
\captionsetup{justification=centering}
\caption{\textcolor{black}{Results on different code graph representations}}
\label{fig:representations}
\end{figure}

\subsection{Ablation Study (RQ4)}

We first investigated the use of AST, PDG, AST+CFG, and AST+PDG as different code graph representations, 
aiming to check the effectiveness of using CPG. As shown \textcolor{black}{Figure}~\ref{fig:representations}, 
the model trained by AST-based graphs generates the worst results, 
meaning the sole syntax information is insufficient when detecting the code's vulnerabilities. 
In general, using heterogeneous code graph representations such as AST+CFG and AST+PDG produces better results than using AST and PDG separately, 
and CPG achieves the best results among all as it combines properties of AST, CFG and PDG.

\begin{table*}[t!]
\caption{Ablation study on graph learning and non-graph learning on CPG} 
\label{tab:hete}
\scalebox{0.9}{
\begin{tabular}{|ll|l|l|l|l|}
\hline
\multicolumn{2}{|l|}{\diagbox[innerwidth=4.9cm]{Model}{Metric}}                                      & ACCURACY & PRECISION & RECALL & F1   \\ \hline
\multicolumn{1}{|l|}{w/o graph-based learning}               & BiLSTM     & 0.72     & 0.74      & 0.65   & 0.69 \\ \hline
\multicolumn{1}{|l|}{\multirow{2}{*}{graph-based learning}} & $HGT_{homo}$  & 0.79     & 0.80      & 0.68   & 0.74 \\ \cline{2-6}
\multicolumn{1}{|l|}{}                                     & $HGT_{heter}$ & \textbf{0.83}     & \textbf{0.92}      & \textbf{0.79}   & \textbf{0.86} \\ \hline
\end{tabular}
}
\end{table*}
Additionally, we explored the importance of heterogeneous graph learning when it comes to encoding the CPG (Table~\ref{tab:hete}). 
As pointed out in the previous section, the core component of enabling heterogeneous graph learning is HGT, 
which presents good performance when incorporating the 
large number neighbors of different types, 
\textit{i.e.}, complex node and edge heterogeneity in the graph structure~\cite{hu2020heterogeneous}. 
To answer this, we implemented HGT$_{homo}$, which is the variant of HGT and only maintains 
a single set of parameters for all relations (e.g., $Q{\text{-}}Linear$, $K{\text{-}}Linear$ and $V{\text{-}}Linear$ matrices share same parameters regardless of types). 
By doing so, HGT$_{homo}$ preserves the graph learning ability while ignoring the various node/edge types in the CPG. 
It can be seen clearly that HGT$_{heter}$ demonstrates a significant performance gain across all metrics, 
proving the great importance of encoding the structural heterogeneity of the program graphs. 
Specifically, HGT$_{heter}$ keeps distinct edge-based metrics $W^{ATT}_{\phi(e)}$ and $W^{MSG}_{\phi(e)}$ for each edge type $\phi(e)$, 
which enables the model to distinguish the semantic differences of relations even between the same node types. 
Additionally, the edge-driven attention mechanism allows the target node to measure the importance of each connected source node with different edge types. 
For instance, in the case of buffer overflow vulnerability in CWE-119, 
the falsely converted \textit{unsign} length variable (source node) 
by \textit{atoi} contributes more to the \textit{memcpy} method (target node) with a larger attention value over the data dependency edge type. 
The importance of this \textit{unsign} variable's node embedding will eventually reflect in the graph-level embedding through readout operation, 
which then helps downstream vulnerability detection task.
Apart from that, we also study the role of graph learning in the experiment. 
Similar to ~\cite{sysevr,lin}, we use BiLSTM as the alternative to encode the code snippets, 
which tokenizes the code representation (i.e., CPG in this case) as input while ignoring the syntax, semantics, and flow structure information in the graphs. The experimental results show that HGT$_{heter}$ outperforms BiLSTM with a greater 10\% performance gain, manifesting the significance of incorporating structure information when modelling the code snippets.

\begin{table}[]

\centering
\caption{The removal of specific edge type and its impact on performance}
\label{otherTo}
\scalebox{0.8}{\begin{tabular}{|l|l|l|l|l|}
\hline
\multirow{2}{*}{\diagbox[]{Edge Type}{Metirc}}  & \multirow{2}{*}{Accuracy} & \multirow{2}{*}{Precision} & \multirow{2}{*}{Recall} & \multirow{2}{*}{F1} \\
                            &                           &                            &                         &                     \\ \hline
w/o Edge Removal                       & \textbf{0.83}                      & 0.80                       & 0.72                    & 0.76                \\ \hline                             
SOURCE\_FILE                & 0.79                      & 0.76                       & 0.70                    & 0.73                \\ \hline
ALIAS\_OF                   & 0.81                      & 0.78                       & 0.72                    & 0.75                \\ \hline
BINDS\_TO                   & 0.78                      & 0.77                       & 0.68                    & 0.72                \\ \hline
INHERITS\_FROM              & 0.82                      & 0.77                       & 0.72                    & 0.74                \\ \hline
AST                         & 0.56                      & 0.60                       & 0.36                    & 0.45                \\ \hline
CONDITION                   & 0.76                      & 0.75                       & 0.66                    & 0.70                \\ \hline
ARGUMENT                    & 0.76                      & 0.67                       & 0.78                    & 0.72                \\ \hline
CALL                        & 0.71                      & 0.68                       & 0.61                    & 0.64                \\ \hline
RECEIVER                    & 0.80                      & 0.79                       & 0.71                    & 0.75                \\ \hline
CFG                         & 0.64                      & 0.80                       & 0.43                    & 0.56                \\ \hline
DOMINATE                    & 0.73                      & 0.74                       & 0.58                    & 0.65                \\ \hline
POST\_DOMINATE              & 0.70                      & 0.83                       & 0.56                    & 0.67                \\ \hline
CDG                         & 0.66                      & 0.68                       & 0.52                    & 0.59                \\ \hline
REACHING\_DEF               & 0.76                      & 0.72                       & 0.76                    & 0.74                \\ \hline
CONTAINS                    & 0.75                      & 0.77                       & 0.64                    & 0.70                \\ \hline
EVAL\_TYPE                  & 0.77                      & 0.76                       & 0.66                    & 0.71                \\ \hline
PARAMETER\_LINK             & 0.80                      & 0.74                       & 0.69                    & 0.71                \\ \hline
TAGGED\_BY                  & 0.80                      & 0.76                       & 0.72                    & 0.74                \\ \hline
BINDS                       & 0.78                      & 0.77                       & 0.67                    & 0.72                \\ \hline
REF                         & 0.79                      & \textbf{0.84}                       & \textbf{0.78}                    & \textbf{0.81}                \\ \hline 

\end{tabular}}

\end{table}

To better examine the impact of each edge type of CPG on the vulnerability detection performance, we remove one edge type at a time on the open-source real-world project called Wireshark. The experimental results are shown in Table~\ref{otherTo}. From the experimental results’ perspective, it is evident that removing most edge types results in a performance drop, indicating that most edges are designed to represent specific program behavior in CPG in the vulnerability detection task. Particularly, the removal of AST edges, CFG edges, and CDG edges leads to a significantly negative impact on the performance. As illustrated in Table~\ref{tab:edge}, AST edges represent the syntax structure of a program, while CFG and CDG edges indicate logical control flow and control dependency from one node to another. Removing these edges thus disrupts the foundational structure of the CPG. However, the deletion of ALIAS\_OF, BINDS\_TO, INHERITS\_FROM, RECEIVER, PARAMETER\_LINK and TAGGED\_BY edges, on the other hand, is less detrimental to the overall performance even though these edges represent the variable/data dependency to some extent. The rationale is that the data dependency carried by these edges is remedied by edges such as CALL, CDG and REACHING\_DEF. In the context of a Code Property Graph (CPG), RECEIVER edges pertain to situations where an object or variable obtains a return value from a function or method call. However, this relationship can be effectively represented by combining REACHING\_DEF and CALL edges. Notably, the elimination of REF edges can lead to performance enhancements. This is due to the fact that REF edges redundantly establish a direct connection between the usage of a variable and its declaration or definition in the program. Such redundancy is already captured by the REACHING\_DEF edges, which comprehensively depict the flow of data encompassing both variable definition and usage.

\subsection{Results on Real-world Open Source Projects (RQ5)}

\begin{table}[]
\centering
\caption{Statistics about FFmpeg, QEMU and Wireshark}
\label{realworldstatstics}
\scalebox{0.8}{\begin{tabular}{|c|cc|c|c|c|}
\hline
\multirow{2}{*}{Dataset} & \multicolumn{2}{c|}{Code Samples}                                & \multirow{2}{*}{Lines of Code} & \multirow{2}{*}{Number of Nodes} & \multirow{2}{*}{Number of Edges} \\ \cline{2-3}
                         & \multicolumn{1}{c|}{Vulnerable Samples} & Non-vulnerable Samples &                                &                                  &                                  \\ \hline
FFmpeg                   & \multicolumn{1}{c|}{4981}               & 4788                   & 721,446                        & 7,906,245                        & 48,731,964                       \\ \hline
QEMU                     & \multicolumn{1}{c|}{7479}               & 10070                  & 875,605                        & 5,276,292                        & 32,376,005                       \\ \hline
Wireshark                & \multicolumn{1}{c|}{637}                   &2635                        & 812,315                        & 2,906,822                        & 22,393,416                       \\ \hline
\end{tabular}
}
\vspace{-3mm}
\end{table}

We also verified the effectiveness of our proposed framework and other baselines 
on real-world programs extracted from C/C++ open-source projects \textbf{\textit{FFmpeg}}, \textbf{\textit{Wireshark}} and \textbf{\textit{QEMU}}. 
The detailed statistics of these projects 
are shown in Table~\ref{realworldstatstics}.

\begin{table*}[t!]
\caption{Results of the comparison with different baselines on real-world projects}
\label{tab:realworld}
\scalebox{1}{
\begin{tabular}{|ll|l|l|l|l||l|}
\hline
\multicolumn{2}{|l|}{\diagbox[innerwidth=4cm]{Dataset}{Metric}{Model}}                & DEVIGN & LIN et al. & DeepWukong & FUNDED & DSHGT \\ \hline
\multicolumn{1}{|l|}{\multirow{4}{*}{FFmpeg}} & ACCURACY  & 0.74   & 0.64       & 0.75       & 0.75   & \textbf{0.79} \\ \cline{2-7} 
\multicolumn{1}{|l|}{}                        & PRECISION & 0.81   & 0.72       & 0.79       & 0.83   & \textbf{0.86} \\ \cline{2-7} 
\multicolumn{1}{|l|}{}                        & RECALL    & 0.68   & 0.58       & 0.71       & 0.67   & \textbf{0.73} \\ \cline{2-7} 
\multicolumn{1}{|l|}{}                        & F1        & 0.70   & 0.66       & 0.78       & 0.79   & \textbf{0.84} \\ \hline
\multicolumn{1}{|l|}{\multirow{4}{*}{QEMU}}   & ACCURACY  & 0.73   & 0.60       & 0.72       & 0.72   & \textbf{0.78} \\ \cline{2-7} 
\multicolumn{1}{|l|}{}                        & PRECISION & 0.78   & 0.70       & 0.80       & 0.76   & \textbf{0.84} \\ \cline{2-7} 
\multicolumn{1}{|l|}{}                        & RECALL    & 0.69   & 0.54       & 0.69       & 0.68   & \textbf{0.73} \\ \cline{2-7} 
\multicolumn{1}{|l|}{}                        & F1        & 0.74   & 0.64       & 0.75       & 0.73   & \textbf{0.82} \\ \hline
\multicolumn{1}{|l|}{\multirow{4}{*}{WireShark}}   & ACCURACY  & 0.77   & 0.71       & 0.75       & 0.79   & \textbf{0.83} \\ \cline{2-7} 
\multicolumn{1}{|l|}{}                        & PRECISION & 0.64   & 0.61       & 0.72       & 0.76   & \textbf{0.80} \\ \cline{2-7} 
\multicolumn{1}{|l|}{}                        & RECALL    & 0.68   & 0.53       & 0.69       & 0.67   & \textbf{0.72} \\ \cline{2-7} 
\multicolumn{1}{|l|}{}                        & F1        & 0.66   & 0.57       & 0.70       & 0.71   & \textbf{0.76} \\ \hline
\end{tabular}
}
\end{table*}
Table~\ref{tab:realworld} records the experimental results of these projects. In general, we observe a drop in performance for both our proposed framework and other baselines on real-world open source projects, as there are fewer labelled vulnerable samples in these three projects than synthetic code samples in \textbf{SARD}. 
This is attributed to the existence of vulnerability label noise in FFmpeg, QEMU and Wireshark since the function-level vulnerabilities are labelled based
on determining vulnerability-fix commits or non-vulnerability fix commits of
these projects. Additionally, Lin et al. underperform for all metrics, especially for \textbf{Recall} (high volumes of false-negative predictions), 
suggesting that only using code tokens in the program to train the detection model is insufficient nor applicable to real-world projects. 
Other baselines present results with marginal differences, while \tool~outperforms all of them. 
Our investigation reveals a substantial presence of vulnerabilities, predominantly Memory Leak and Buffer Overflow, within these projects. To enhance the detection outcomes, a refined representation of the information associated with various edge types, specifically control flow and dependency information, is imperative in the graph-level embedding process. 
Since \tool~learns on the CPG without losing the heterogeneity attributes, thus is more generalized even on real-world projects, especially in Wireshark. Overall, \tool~delivers the best performance for \textbf{Accuracy} (7.8\% improvement on average), \textbf{Precision} (9\% improvement on average), \textbf{Recall} (7.6\% improvement on average) and \textbf{F1} (10.4\% improvement on average). A promising way to improve the performance on real-world projects is to adapt state-of-the-art zero-shot or few-shot learning specifically for heterogeneous graph networks.

\begin{table*}[t!]
\caption{Experimental results on unseen vulnerability types in Wireshark}
\label{tab:maskVul}
\scalebox{0.8}{
\begin{tabular}{|c|l|cccc|}
\hline
\multirow{2}{*}{\begin{tabular}[c]{@{}c@{}}Proportion of  Masked \\ Vulnerability Types\end{tabular}} & \multicolumn{1}{c|}{\multirow{2}{*}{Masked Vulnerability Types}}                                                                                                                                                                                                                                        & \multicolumn{4}{c|}{Evaluation Metrics}                                                                    \\ \cline{3-6} 
                                                                                                      & \multicolumn{1}{c|}{}                                                                                                                                                                                                                                                                                   & \multicolumn{1}{c|}{Accuracy} & \multicolumn{1}{c|}{Precision} & \multicolumn{1}{c|}{Recall} & F1   \\ \hline
0\%                                                                                                   & \multicolumn{1}{c|}{-}                                                                                                                                                                                                                                                                                  & \multicolumn{1}{c|}{\textbf{0.83}}     & \multicolumn{1}{c|}{\textbf{0.80}}      & \multicolumn{1}{c|}{\textbf{0.72}}   & \textbf{0.76} \\ \hline
10\%                                                                                                  & CWE-416, CWE-820, CWE-414, CWE-543, CWE-195                                                                                                                                                                                                                                                             & \multicolumn{1}{c|}{0.80}     & \multicolumn{1}{c|}{0.78}      & \multicolumn{1}{c|}{0.68}   & 0.73 \\ \hline
20\%                                                                                                  & \begin{tabular}[c]{@{}l@{}}CWE-805, CWE-663, CWE-843, CWE-820, CWE-590,\\ CWE-785, CWE-833, CWE-839, CWE-170, CWE-821\end{tabular}                                                                                                                                                                      & \multicolumn{1}{c|}{0.76}     & \multicolumn{1}{c|}{0.74}      & \multicolumn{1}{c|}{0.63}   & 0.68 \\ \hline
30\%                                                                                                  & \begin{tabular}[c]{@{}l@{}}CWE-414, CWE-190, CWE-771, CWE-828, CWE-834,\\ CWE-191, CWE-412, CWE-764, CWE-401, CWE-369,\\ CWE-416, CWE-367, CWE-129, CWE-170, CWE-789,\\ CWE-459\end{tabular}                                                                                                            & \multicolumn{1}{c|}{0.71}     & \multicolumn{1}{c|}{0.71}      & \multicolumn{1}{c|}{0.56}   & 0.63 \\ \hline
40\%                                                                                                  & \begin{tabular}[c]{@{}l@{}}CWE-839, CWE-120, CWE-543, CWE-775, CWE-663, \\ CWE-806, CWE-459, CWE-412, CWE-415, CWE-170,\\ CWE-400, CWE-196, CWE-674, CWE-363, CWE-773, \\ CWE-824, CWE-820, CWE-414, CWE-401, CWE-761, \\ CWE-805\end{tabular}                                                          & \multicolumn{1}{c|}{0.68}     & \multicolumn{1}{c|}{0.70}      & \multicolumn{1}{c|}{0.52}   & 0.60 \\ \hline
50\%                                                                                                  & \begin{tabular}[c]{@{}l@{}}CWE-543, CWE-820, CWE-843, CWE-414, CWE-415,\\ CWE-479, CWE-412, CWE-363, CWE-459, CWE-367, \\ CWE-834, CWE-126, CWE-129, CWE-663, CWE-775, \\ CWE-674, CWE-170, CWE-831, CWE-764, CWE-828, \\ CWE-821, CWE-416, CWE-127, CWE-590, CWE-401, \\ CWE-190, CWE-839\end{tabular} & \multicolumn{1}{c|}{0.64}     & \multicolumn{1}{c|}{0.66}      & \multicolumn{1}{c|}{0.45}   & 0.54 \\ \hline
\end{tabular}
}
\end{table*}

We also conducted the experiment to verify the performance of \textit{DSHGT} in unseen vulnerabilities in real-world projects (Table~\ref{tab:maskVul}). Specifically, we randomly mask proportional vulnerability types (from 0\% up to 50\% of the total 54 uniquely different vulnerability types)  in \textbf{\textit{Wireshark}} in a non-overlapping manner to emulate the unseen vulnerability scenarios. These masked vulnerability types are then used as the test set so that the impact on performance can be observed directly. To encapsulate the findings, our analysis presents a clear inverse correlation between the extent of vulnerability type obfuscation and the vulnerability detection performance of the model. As shown in Table~\ref{tab:maskVul}, there is a marked depreciation in all metrics, which descends from an initial accuracy of 83\% (0\% vulnerability masked) to a diminished value of 64\% (50\% vulnerability masked). Concurrently, the F1 score also experiences a similar degeneration, tapering from 76\% to 54\%. This finding reveals that the model exhibits certain generalization capacities, sustaining a consistent predictive performance despite deliberately masking a sheer proportion of vulnerabilities.

\section{Threat to Validity}
\label{sec:validity}

\textbf{External validity}:
In our experiments, we chose 4 baselines: 
\textbf{LIN \textit{et al.}}~\cite{lin},
\textbf{DEVIGN}~\cite{devign}, 
\textbf{FUNDED}~\cite{funded} 
and 
\textbf{DeepWukong}~\cite{deepwukong}. 
We believe these 4 baselines could represent the state-of-the-art research outputs, 
including non-graph learning methods for code tokens and graph learning methods for code graph representations.
For the datasets used, we used 22 categories of C/C++, Java and PHP from \textbf{SARD}, 
which shall cover the most commonly seen vulnerability types.
Due to the lack of labeled vulnerabilities datasets from real-world projects, 
we could only evaluate \tool~on the real-world open-source projects 
(\textit{i.e.} \textbf{\textit{FFmpeg}}, \textbf{\textit{QEMU}} and \textbf{\textit{WireShark}}, 
some of which have also been used by~\cite{li2021vuldeelocator},~\cite{lin} and~\cite{devign}).
As these three projects also allow code contributions, 
we believe they are good representations of state-of-the-art real-world projects. 
Even though our proposed \tool~has shown the transferability to detect Java and PHP vulnerabilities after being trained on the C/C++ dataset, 
the vulnerability datasets for programming languages other than C/C++ are still limited. 
We will continue to work on this issue and build up related datasets for future usage in this community.

\noindent\textbf{Internal validity}:
\revised{The quality of code comments used for training might not align with the ground truth perfectly due to the semantic complexity of codes and human errors.
As discussed in Section~\ref{sec:rq3}, the hand-written code comment highly relies on the programmers' personal habits,
and different programmers could provide different code comments for the same source code segment.
Thus, we assign $\lambda$ to balance the contribution from the code comment supervisor, and our empirical study shows that our \tool~reaches the highest performance when $\lambda = 0.2$. 
We plan to use \tool~in the future for those real-world software projects where 
we have finer-granularity control of the code comment quality to remove confounding variables such as human errors or other noises.}

\noindent\textbf{Construct validity}
The layer depth of \textit{HGT} in \textit{DSHGT} is set to 3 based on the empirical study result (as shown in Section~\ref{sec:rq3}). 
The target node embedding is updated through the attentive message passing from neighbor nodes.
As the layer numbers increase, the \textit{HGT} will eventually face the oversmoothing problem~\cite{chen2020measuring}, 
where distinct information conveyed by neighbors becomes hard to distinguish owing to the over-iterative message passing and leads to performance degradation.
We leverage a simple 1-layer LSTM as the decoder for the \textcolor{black}{code comment} supervisor part, 
which already demonstrates promising results in the experiment. 
It suggests the potential of using auxiliary \textcolor{black}{code comments} as supplementary information to assist the vulnerability detection task. 
As a future task, we can leverage more calibrated semantic extraction models 
such as those pretrained NLP models~\cite{kalyan2021ammus} as the \textcolor{black}{code comment} decoder.

\section{Conclusion}
\label{sec:conclusion}

In this paper, we present our pioneer study of using heterogeneous graph learning for detecting software vulnerabilities. 
In this work, we have chosen to use CPG as the heterogeneous graph and customized a dual-supervisor graph learning model. 
The extensive experiments on synthetic and real-world software projects show promising results. 
We are one of the first to explore the importance of conducting research in heterogeneous graph learning 
to detect software vulnerability for the software engineering community. 
Besides good results in different programming languages and real-world projects, 
software engineers can leverage the largely open-sourced new results from graph learning 
and NLP communities to further enhance this line of research to have better heterogeneous graph representation other than CPG, 
leverage more robust learning models for latent features in the underlying heterogeneous graph, 
and utilize more reliable and controllable supervisors in addition to the binary vulnerability oracle. \revise{In our future research, we plan to explore the feasibility of integrating original code tokens with the code graph structure during the learning process. We believe that this integration has the potential to address these outlier scenarios and enhance the overall performance of vulnerability detection.}


\bibliographystyle{ACM-Reference-Format}
\bibliography{ref}







\clearpage
\onecolumn

\section{Appendix}

\label{sec:app}
We show supplementary tables in this Appendix.

\begin{table}[h!]
\centering
\caption{All node types in CPG.}
\label{tab:node}
\scalebox{0.55}{
\begin{tabular}{|l|l|}
\hline
Node\_type                    & Description                                                                                                                     \\ \hline
META\_DATA                    & This node contains the CPG meta data.                                                                                           \\ \hline
FILE                          & File nodes represent source files or a shared objects from which the CPG was generated.                                         \\ \hline
NAMESPACE                     & This node represents a namespace.                                                                                               \\ \hline
NAMESPACE\_BLOCK              & A reference to a namespace.                                                                                                     \\ \hline
METHOD                        & This node represents procedures, functions, methods.                                                                            \\ \hline
METHOD\_PARAMETER\_IN         & This node represents a formal input parameter.                                                                                  \\ \hline
METHOD\_PARAMETER\_OUT        & This node represents a formal output parameter.                                                                                 \\ \hline
METHOD\_RETURN                & This node represents an (unnamed) formal method return parameter.                                                               \\ \hline
MEMBER                        & This node represents a type member of a class, struct or union.                                                                 \\ \hline
TYPE                          & This node represents a type instance, a concrete instantiation of a type declaration.                                           \\ \hline
TYPE\_ARGUMENT                & This node represents a type argument which is used to instantiate a parametrized type.                                          \\ \hline
TYPE\_DECL                    & This node represents a type declaration.                                                                                        \\ \hline
TYPE\_PARAMETER               & This node represents a formal type parameter,  the type parameter as given in a type-parametrized method or type declaration.   \\ \hline
AST\_NODE                     & This is the base type for all nodes of the abstract syntax tree (AST).                                                          \\ \hline
BLOCK                         & This node represents a compound statement.                                                                                      \\ \hline
CALL                          & A (function/method/procedure) call.                                                                                             \\ \hline
CALL\_REPR                    & This is the base class of  CALL.                                                                                                \\ \hline
CONTROL\_STRUCTURE            & This node represents a control structure as introduced by control structure statements.                                         \\ \hline
EXPRESSION                    & This node is the base class for all nodes that represent code pieces that can be evaluated.                                     \\ \hline
FIELD\_IDENTIFIER             & This node represents the field accessed in a field access.                                                                      \\ \hline
IDENTIFIER                    & This node represents an identifier as used when referring to a variable by name..                                               \\ \hline
JUMP\_LABEL                   & This node specifies the label and thus the JUMP\_TARGET of control structures BREAK and CONTINUE.                               \\ \hline
JUMP\_TARGET                  & A jump target is a location that has been specifically marked as the target of a jump.                                          \\ \hline
LITERAL                       & This node represents a literal such as an integer or string constant.                                                           \\ \hline
LOCAL                         & This node represents a local variable.                                                                                          \\ \hline
METHOD\_REF                   & This node represents a reference to a method/function/procedure as it appears when a method is passed as an argument in a call. \\ \hline
MODIFIER                      & This node represents a (language-dependent) modifier such as `static', `private' or `public'.                                   \\ \hline
RETURN                        & This node represents a return instruction.                                                                                      \\ \hline
TYPE\_REF                     & Reference to a type/class.                                                                                                      \\ \hline
UNKNOWN                       & This node represents an AST node but there is no suitable AST node type.                                                        \\ \hline
CFG\_NODE                     & Base class for all control flow nodes.                                                                                          \\ \hline
COMMENT                       & A source code comment node.                                                                                                     \\ \hline
FINDING                       & Finding nodes are used to store analysis results in the graph.                                                                  \\ \hline
KEY\_VALUE\_PAIR              & This node represents a key-value pair.                                                                                          \\ \hline
LOCATION                      & A location node summarizes a source code location.                                                                              \\ \hline
TAG                           & This node represents a tag.                                                                                                     \\ \hline
TAG\_NODE\_PAIR               & This node contains an arbitrary node and an associated tag node.                                                                \\ \hline
CONFIG\_FILE                  & This node represent a configuration file.                                                                                       \\ \hline
BINDING                       & BINDING nodes represent name-signature pairs that can be resolved at a type declaration.                                        \\ \hline
ANNOTATION                    & A method annotation node.                                                                                                       \\ \hline
ANNOTATION\_LITERAL           & A literal value assigned to an ANNOTATION\_PARAMETER.                                                                           \\ \hline
ANNOTATION\_PARAMETER         & Formal annotation parameter.                                                                                                    \\ \hline
ANNOTATION\_PARAMETER\_ASSIGN & Assignment of annotation argument to annotation parameter.                                                                      \\ \hline
ARRAY\_INITIALIZER            & Initialization construct for arrays.                                                                                            \\ \hline
DECLARATION                   & Base node class for all declarations.                                                                                           \\ \hline
\end{tabular}
}
\end{table}

\begin{table}[h!]
\centering
\caption{All edge types in CPG.}
\label{tab:edge}
\scalebox{0.55}{
\begin{tabular}{|l|l|}
\hline
Edge\_type      & Description                                                                                                                                        \\ \hline
SOURCE\_FILE    & This edge connects a node to the node that represents its source file.                                                                             \\ \hline
ALIAS\_OF       & This edge represents an alias relation between a type declaration and a type.                                                                      \\ \hline
BINDS\_TO       & This edge connects type arguments to type parameters to indicate that the type argument is used to instantiate the type parameter.                 \\ \hline
INHERITS\_FROM  & Inheritance relation between a type declaration and a type.                                                                                        \\ \hline
AST             & This edge connects a parent node to its child in the abstract syntax tree.                                                                         \\ \hline
CONDITION       & The edge connects control structure nodes to the expressions that holds their conditions.                                                          \\ \hline
ARGUMENT        & Argument edges connect call sites to their arguments as well as the expressions that return.                                                       \\ \hline
CALL            & This edge connects call sites.                                                                                                                     \\ \hline
RECEIVER        & RECEIVER edges connect call sites to their receiver arguments.                                                                                     \\ \hline
CFG             & This edge indicates control flow from the source to the destination node.                                                                          \\ \hline
DOMINATE        & This edge indicates that the source node immediately dominates the destination node.                                                               \\ \hline
POST\_DOMINATE  & This edge indicates that the source node immediately post dominates the destination node.                                                          \\ \hline
CDG             & A CDG edge expresses that the destination node is control dependent on the source node.                                                            \\ \hline
REACHING\_DEF   & A reaching definition edge indicates that a variable produced at the source node reaches the destination node without being reassigned on the way. \\ \hline
CONTAINS        & This edge connects a node to the method that contains it.                                                                                          \\ \hline
EVAL\_TYPE      & This edge connects a node to its evaluation type.                                                                                                  \\ \hline
PARAMETER\_LINK & This edge connects a method input parameter to the corresponding method output parameter.                                                          \\ \hline
TAGGED\_BY      & This edge connects a node to the a TAG node.                                                                                                       \\ \hline
BINDS           & This edge connects a type declaration (TYPE\_DECL) with a binding node (BINDING).                                                                  \\ \hline
REF             & This edge indicates that the source node is an identifier that denotes access to the destination node.                                             \\ \hline
\end{tabular}
}
\end{table}

\begin{table}[t!]
\centering
\caption{Dataset overview}
\label{tab:dataset}
\scalebox{0.6}{
\begin{tabular}{|l|l|l|l|l|}
\hline
Language                                          & Vul Category & Description                                                                                                                                                                                                                                & Vulnerable Samples & Non-vulnerable Samples \\ \hline
\multirow{17}{*}{C/C++}                           & CWE-119      & Improper Restriction of Operations within the Bounds of a Memory Buffer                                                                                                                                                                    & 1597               & 5406                   \\ \cline{2-5} 
                                                  & CWE-400      & Uncontrolled Resource Consumption                                                                                                                                                                                                          & 2786               & 8563                   \\ \cline{2-5} 
                                                  & CWE-404      & Improper Resource Shutdown or Release                                                                                                                                                                                                      & 1054               & 3748                   \\ \cline{2-5} 
                                                  & CWE-369      & Divide By Zero                                                                                                                                       & 2691               & 8266                   \\ \cline{2-5} 
                                                  & CWE-125      & Out-of-bounds Read                                                                                                      & 1843               & 6310                   \\ \cline{2-5}
                                                  & CWE-416      & Use After Free                                                                                                                                                                                                                             & 1372               & 4439                   \\ \cline{2-5}
                                                  & CWE-191      & Integer Underflow (Wrap or Wraparound)                                                                                                                                                                                                     & 3473               & 10572                  \\ \cline{2-5} 
                                                  & CWE-476      & NULL Pointer Dereference                                                                                                                                                                                                                   & 1467               & 7336                   \\ \cline{2-5} 
                                                  &CWE-20        &Improper Input Validation       &2617    &5718    \\ \cline{2-5} 
                                                  & CWE-78       & \begin{tabular}[c]{@{}l@{}}Improper Neutralization of Special Elements used in an OS Command\\  (`OS Command Injection')\end{tabular}                                                                                                      & 10519              & 37854                  \\ \cline{2-5} 
                                                  &CWE-79        &\begin{tabular}[c]{@{}l@{}}Improper Neutralization of Input During Web Page Generation\\   ('Cross-site Scripting')\end{tabular}       &1637    &5275    \\ \cline{2-5}
                                                  &CWE-89        &\begin{tabular}[c]{@{}l@{}}Improper Neutralization of Special Elements used in an SQL Command \\   ('SQL Injection')\end{tabular}       &2164    &5796    \\ \cline{2-5}
                                                  & CWE-772      & Missing Release of Resource after Effective Lifetime                                                                                                                                                                                       & 4755               & 11847                  \\ \cline{2-5} 
                                                  & CWE-467      & Use of sizeof(~) on a Pointer Type                                                                                                                                                                                                          & 867                & 3573                   \\ \cline{2-5} 
                                                  & CWE-190      & Integer Overflow or Wraparound                                                                                                                                                                                                             & 3648               & 10487                  \\ \cline{2-5} 
                                                  & CWE-770      & Allocation of Resources Without Limits or Throttling                                                                                                                                                                                       & 843                & 2973                   \\ \cline{2-5} 
                                                  & CWE-666      & Operation on Resource in Wrong Phase of Lifetime                                                                                                                                                                                           & 1321               & 4589                   \\ \cline{2-5} 
                                                  & CWE-665      & Improper Initialization                                                                                                                                                                                                                    & 1436               & 4680                   \\ \cline{2-5} 
                                                  & CWE-758      & Reliance on Undefined, Unspecified, or Implementation-Defined Behavior                                                                                                                                                                     & 1674               & 5745                   \\ \cline{2-5} 
                                                  & CWE-469      & Use of Pointer Subtraction to Determine Size                                                                                                                                                                                               & 667                & 2745                   \\ \cline{2-5} 
                                                  & CWE-676      & Use of Potentially Dangerous Function                                                                                                                                                                                                      & 478                & 1394                   \\ \cline{2-5} 
                                                  & CWE-834      & Excessive Iteration                                                                                                                                                                                                                        & 362                & 1644                   \\ \hline
\end{tabular}
}
\end{table}

\begin{table}[]

\centering
\caption{Supplementary comparison with different baselines on SARD (Precision and Recall)}
\label{otherTwo}
\scalebox{0.7}{\begin{tabular}{|l|ll|ll|ll|ll||ll|ll|}
\hline
\multirow{3}{*}{\diagbox[]{DataSet}{Metric}{Model}} & \multicolumn{2}{l|}{\multirow{2}{*}{DEVIGN}} & \multicolumn{2}{l|}{\multirow{2}{*}{LIN et al.}} & \multicolumn{2}{l|}{\multirow{2}{*}{FUNDED}} & \multicolumn{2}{l||}{\multirow{2}{*}{DeepWukong}} & \multicolumn{2}{l|}{\multirow{2}{*}{DSHGTnoAnno}} & \multicolumn{2}{l|}{\multirow{2}{*}{DSHGT}} \\
                        & \multicolumn{2}{l|}{}                        & \multicolumn{2}{l|}{}                            & \multicolumn{2}{l|}{}                        & \multicolumn{2}{l||}{}                            & \multicolumn{2}{l|}{}                             & \multicolumn{2}{l|}{}                       \\ \cline{2-13} 
                        & \multicolumn{1}{l|}{Precision}    & Recall   & \multicolumn{1}{l|}{Precision}      & Recall     & \multicolumn{1}{l|}{Precision}    & Recall   & \multicolumn{1}{l|}{Precision}      & Recall     & \multicolumn{1}{l|}{Precision}      & Recall      & \multicolumn{1}{l|}{Precision}   & Recall   \\ \hline
CWE-119                 & \multicolumn{1}{l|}{0.85}         & 0.77     & \multicolumn{1}{l|}{0.75}           & 0.84       & \multicolumn{1}{l|}{0.82}         & 0.90     & \multicolumn{1}{l|}{0.88}           & 0.80       & \multicolumn{1}{l|}{0.89}           & 0.83        & \multicolumn{1}{l|}{\textbf{0.92}}        & \textbf{0.83}     \\ \hline
CWE-400                 & \multicolumn{1}{l|}{0.80}         & 0.82     & \multicolumn{1}{l|}{0.82}           & 0.71       & \multicolumn{1}{l|}{0.86}         & 0.75     & \multicolumn{1}{l|}{0.82}           & 0.76       & \multicolumn{1}{l|}{0.84}           & 0.82        & \multicolumn{1}{l|}{\textbf{0.87}}        & \textbf{0.83}     \\ \hline
CWE-404                 & \multicolumn{1}{l|}{0.86}         & 0.78     & \multicolumn{1}{l|}{0.72}           & 0.76       & \multicolumn{1}{l|}{0.86}         & 0.84     & \multicolumn{1}{l|}{0.87}           & 0.75       & \multicolumn{1}{l|}{0.86}           & 0.90        & \multicolumn{1}{l|}{\textbf{0.89}}        & \textbf{0.91}     \\ \hline
CWE-369                 & \multicolumn{1}{l|}{0.84}         & 0.73     & \multicolumn{1}{l|}{0.77}           & 0.83       & \multicolumn{1}{l|}{0.90}         & 0.79     & \multicolumn{1}{l|}{0.90}           & 0.84       & \multicolumn{1}{l|}{\textbf{0.93}}           & \textbf{0.89}        & \multicolumn{1}{l|}{0.92}        & 0.84     \\ \hline
CWE-191                 & \multicolumn{1}{l|}{0.83}         & 0.70     & \multicolumn{1}{l|}{0.77}           & 0.69       & \multicolumn{1}{l|}{0.93}         & \textbf{0.87}     & \multicolumn{1}{l|}{0.88}           & 0.75       & \multicolumn{1}{l|}{0.90}           & 0.84        & \multicolumn{1}{l|}{\textbf{0.95}}        & \textbf{0.87}     \\ \hline
CWE-476                 & \multicolumn{1}{l|}{0.90}         & 0.84     & \multicolumn{1}{l|}{0.79}           & 0.87       & \multicolumn{1}{l|}{\textbf{0.93}}         & 0.82     & \multicolumn{1}{l|}{0.82}           & \textbf{0.90}       & \multicolumn{1}{l|}{0.86}           & 0.82        & \multicolumn{1}{l|}{0.92}        & 0.86     \\ \hline
CWE-467                 & \multicolumn{1}{l|}{0.80}         & 0.88     & \multicolumn{1}{l|}{0.77}           & 0.85       & \multicolumn{1}{l|}{0.82}         & 0.91     & \multicolumn{1}{l|}{\textbf{0.83}}           & 0.89       & \multicolumn{1}{l|}{0.79}           & 0.87        & \multicolumn{1}{l|}{0.82}        & \textbf{0.93}     \\ \hline
CWE-78                  & \multicolumn{1}{l|}{0.81}         & 0.87     & \multicolumn{1}{l|}{0.75}           & 0.84       & \multicolumn{1}{l|}{0.82}         & 0.90     & \multicolumn{1}{l|}{\textbf{0.84}}           & 0.82       & \multicolumn{1}{l|}{0.79}           & \textbf{0.94}        & \multicolumn{1}{l|}{0.80}        & 0.88     \\ \hline
CWE-772                 & \multicolumn{1}{l|}{0.83}         & 0.72     & \multicolumn{1}{l|}{0.84}           & 0.78       & \multicolumn{1}{l|}{0.91}         & 0.83     & \multicolumn{1}{l|}{0.88}           & 0.79       & \multicolumn{1}{l|}{0.90}           & \textbf{0.86}        & \multicolumn{1}{l|}{\textbf{0.94}}        & 0.83     \\ \hline
CWE-190                 & \multicolumn{1}{l|}{0.86}         & 0.80     & \multicolumn{1}{l|}{0.83}           & 0.75       & \multicolumn{1}{l|}{0.88}         & 0.80     & \multicolumn{1}{l|}{0.85}           & 0.81       & \multicolumn{1}{l|}{0.84}           & 0.80        & \multicolumn{1}{l|}{\textbf{0.90}}        & \textbf{0.82}     \\ \hline
CWE-770                 & \multicolumn{1}{l|}{0.82}         & 0.86     & \multicolumn{1}{l|}{0.76}           & 0.84       & \multicolumn{1}{l|}{0.85}         & 0.89     & \multicolumn{1}{l|}{\textbf{0.88}}           & 0.86       & \multicolumn{1}{l|}{0.82}           & 0.90        & \multicolumn{1}{l|}{0.87}        & \textbf{0.91}     \\ \hline
CWE-666                 & \multicolumn{1}{l|}{0.89}         & 0.80     & \multicolumn{1}{l|}{0.89}           & 0.83       & \multicolumn{1}{l|}{0.93}         & 0.87     & \multicolumn{1}{l|}{0.94}           & 0.90       & \multicolumn{1}{l|}{0.89}           & \textbf{0.93}        & \multicolumn{1}{l|}{\textbf{0.95}}        & 0.91     \\ \hline
CWE-665                 & \multicolumn{1}{l|}{0.84}         & 0.90     & \multicolumn{1}{l|}{0.75}           & 0.84       & \multicolumn{1}{l|}{0.84}         & 0.92     & \multicolumn{1}{l|}{0.86}           & 0.92       & \multicolumn{1}{l|}{\textbf{0.93}}           & 0.91        & \multicolumn{1}{l|}{0.89}        & \textbf{0.95}     \\ \hline
CWE-758                 & \multicolumn{1}{l|}{0.90}         & 0.84     & \multicolumn{1}{l|}{0.85}           & 0.81       & \multicolumn{1}{l|}{0.85}         & 0.91     & \multicolumn{1}{l|}{0.90}           & 0.94       & \multicolumn{1}{l|}{0.87}           & 0.91        & \multicolumn{1}{l|}{\textbf{0.91}}        & \textbf{0.95}     \\ \hline
CWE-469                 & \multicolumn{1}{l|}{0.85}         & 0.74     & \multicolumn{1}{l|}{0.81}           & 0.72       & \multicolumn{1}{l|}{0.88}         & 0.79     & \multicolumn{1}{l|}{0.84}           & 0.75       & \multicolumn{1}{l|}{0.87}           & 0.81        & \multicolumn{1}{l|}{\textbf{0.89}}        & \textbf{0.83}     \\ \hline
CWE-676                 & \multicolumn{1}{l|}{0.77}         & 0.83     & \multicolumn{1}{l|}{0.72}           & 0.78       & \multicolumn{1}{l|}{0.88}         & 0.94     & \multicolumn{1}{l|}{0.86}           & 0.80       & \multicolumn{1}{l|}{0.84}           & \textbf{0.95}        & \multicolumn{1}{l|}{\textbf{0.87}}        & \textbf{0.95}     \\ \hline
CWE-834                 & \multicolumn{1}{l|}{0.68}         & 0.57     & \multicolumn{1}{l|}{0.70}           & \textbf{0.78}       & \multicolumn{1}{l|}{0.83}         & 0.75     & \multicolumn{1}{l|}{0.80}           & 0.65       & \multicolumn{1}{l|}{0.82}           & 0.71        & \multicolumn{1}{l|}{\textbf{0.86}}        & \textbf{0.78}     \\ \hline
CWE-79                  & \multicolumn{1}{l|}{0.86}         & 0.82     & \multicolumn{1}{l|}{0.88}           & 0.82       & \multicolumn{1}{l|}{0.84}         & 0.82     & \multicolumn{1}{l|}{0.87}           & \textbf{0.89}       & \multicolumn{1}{l|}{0.90}           & 0.84        & \multicolumn{1}{l|}{\textbf{0.92}}        & 0.88     \\ \hline
CWE-89                  & \multicolumn{1}{l|}{0.84}         & 0.80     & \multicolumn{1}{l|}{0.81}           & 0.79       & \multicolumn{1}{l|}{0.88}         & 0.82     & \multicolumn{1}{l|}{0.89}           & 0.80       & \multicolumn{1}{l|}{0.88}           & 0.82        & \multicolumn{1}{l|}{\textbf{0.91}}        & \textbf{0.83}     \\ \hline
CWE-416                 & \multicolumn{1}{l|}{0.83}         & \textbf{0.87}     & \multicolumn{1}{l|}{0.82}           & 0.80       & \multicolumn{1}{l|}{0.90}         & 0.84     & \multicolumn{1}{l|}{0.87}           & 0.81       & \multicolumn{1}{l|}{0.90}           & 0.86        & \multicolumn{1}{l|}{\textbf{0.93}}        & \textbf{0.87}     \\ \hline
CWE-20                  & \multicolumn{1}{l|}{0.91}         & 0.85     & \multicolumn{1}{l|}{0.87}           & 0.79       & \multicolumn{1}{l|}{0.92}         & 0.86     & \multicolumn{1}{l|}{0.91}           & 0.83       & \multicolumn{1}{l|}{0.92}           & 0.88        & \multicolumn{1}{l|}{\textbf{0.93}}        & \textbf{0.91}     \\ \hline
CWE-125                 & \multicolumn{1}{l|}{0.90}         & 0.81     & \multicolumn{1}{l|}{0.87}           & 0.79       & \multicolumn{1}{l|}{0.85}         & \textbf{0.89}     & \multicolumn{1}{l|}{0.88}           & \textbf{0.89}       & \multicolumn{1}{l|}{0.88}           & 0.84        & \multicolumn{1}{l|}{\textbf{0.92}}        & 0.86     \\ \hline
\end{tabular}}

\end{table}

\end{document}